\documentclass[12pt,showpacs,showkeys,preprintnumbers,nofootinbib]{revtex4-1}

\usepackage{amssymb,amsmath,graphics,graphicx}  

\usepackage{rotate,color,colortbl}
\usepackage[dvipsnames]{xcolor}

\usepackage{subcaption}

\usepackage{tikz}
\usetikzlibrary{shapes,arrows}

\begin{document}

\newcommand{\pderiv}[2]{\frac{\partial #1}{\partial #2}}
\newcommand{\deriv}[2]{\frac{d #1}{d #2}}

\title{Sudden transitions in coupled opinion and epidemic dynamics with vaccination}

\author{Marcelo A. Pires $^{1,2}$}
\thanks{piresma@cbpf.br}

\author{Andr\'{e} L. Oestereich $^{2}$}
\thanks{andrelo@id.uff.br}

\author{Nuno Crokidakis $^{2}$}
\thanks{nuno@if.uff.br}

\affiliation{
$^{1}$ Centro Brasileiro de Pesquisas F\'isicas, Rio de Janeiro/RJ, Brazil \\ 
$^{2}$ Instituto de F\'{\i}sica, Universidade Federal Fluminense, Niter\'oi/RJ, Brazil}

\date{\today}

\begin{abstract}
This work consists of an epidemic model with vaccination coupled with an opinion dynamics. Our objective was to study how disease risk perception can influence opinions about vaccination and therefore the spreading of the disease. Differently from previous works we have considered continuous opinions. The epidemic spreading is governed by an SIS-like model with an extra vaccinated state. In our model individuals  vaccinate with a probability proportional to their opinions. The opinions change due to peer influence in pairwise interactions. The epidemic feedback to the opinion dynamics acts as an external field increasing the vaccination probability. We performed Monte Carlo simulations in fully-connected populations. Interestingly we observed the emergence of a first-order phase transition, besides the usual active-absorbing phase transition presented in the SIS model. Our simulations also show that with a certain combination of parameters, an increment in the initial fraction of the population that is pro-vaccine has a twofold effect: it can lead to smaller epidemic outbreaks in the short term, but it also contributes to the survival of the chain of  infections in the long term. Our results also suggest that it is possible that more effective vaccines can decrease the long-term vaccine coverage. This is a counterintuitive outcome, but it is in line with empirical observations that vaccines can become a victim of their own success.
\end{abstract}


\maketitle


\section{Introduction}

Statistical physics is the branch of science that deals with macroscopic phenomena that emerge from the microscopic interactions of its constituent units. Its recognition is increasing in many different research fields such as social, epidemic and vaccination dynamics \cite{Wang2016,Satorras2015,Castellano2009,Galam2012,Sen2013}. The interest of physicists in such systems range from theoretical questions \cite{Moreno2002,Satorras2001,CrokidakisMenezes2012} to practical concerns \cite{Antulov2015,Kitsak2010,Xiong2017}.

In the line of collective phenomena a challenging issue is: what are the possible macroscopic scenarios arising from a coupled vaccination and opinion dynamics? This is  an important topic because the success or failure of a vaccination campaign does not only depends on vaccine-accessibility, vaccine-efficacy and epidemiological variables, but it also depends on the public opinion about vaccination. For instance, in 2010 the French government requested vaccine for H1N1 for 90 million individuals, but about 6 million of the vaccines were effectively used by the population \cite{Galam2010}.  In \cite{Salathe_Bonhoeffer2008} the authors highlight that ``many high-income countries currently experience large outbreaks of vaccine-preventable diseases such as measles despite the availability of highly effective vaccines''. Still in \cite{Salathe_Bonhoeffer2008}  the authors suggest that, as a consequence of the opinion dynamics concerning the vaccination, ``the current estimates of vaccination coverage necessary to avoid outbreaks of vaccine-preventable diseases might be too low''.

Much progress is being achieved about the impact of a dynamical vaccination behavior on disease spreading under a vaccination program, as extensively reviewed in \cite{Wang2016,Wang2015,Verelst_Willem_Beutels2016,Funk2010}. But there is a lack of studies tackling opinions about vaccination as continuous variables. In order to fill this gap, we study the possible emergent scenarios of an epidemic model with vaccination coupled with a social dynamics considering continuous opinions. 
The use of discrete opinions is appropriated to model many situations such as yes/no referendum. However, there are cases as in vaccination dynamics that is more suitable to use continuous opinions since with them: (i) an agent can become surer of his opinion after encountering another one that holds a similar view about the subject (strengthening of opinions); (ii) opinions undergo more gradual changes than theirs discrete counterpart; (iii) it is possible to observe both moderate and extremist opinions in the population. 



\section{Model}

\begin{figure}[h]
\centering
\includegraphics[width=0.8\linewidth]{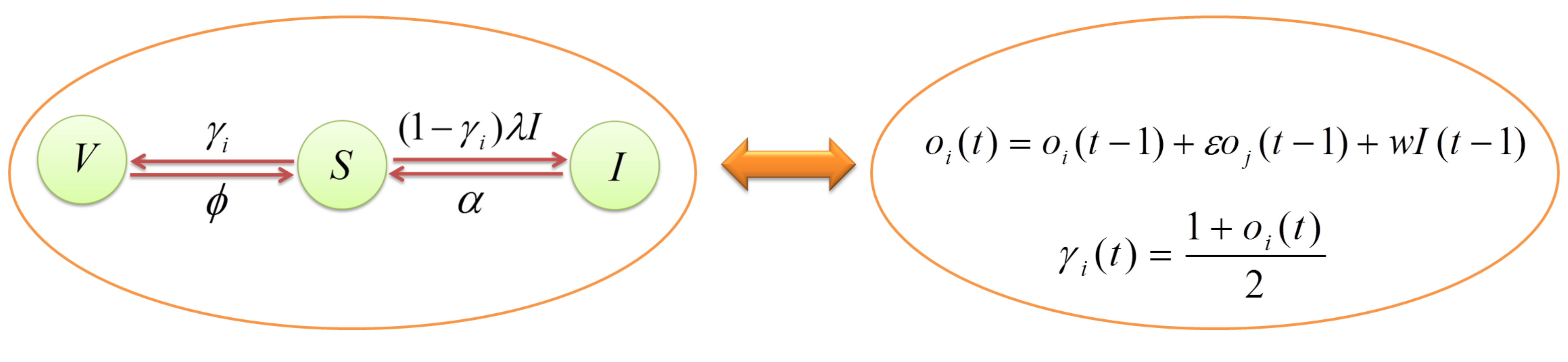}
\caption{Coupled vaccination and  continuous opinion dynamics schematics. The coupling between opinion and epidemic goes in both ways. The opinions influence the vaccination probability, by changing $\gamma_{i}$. The epidemic spreading increases the propensity of an agent getting vaccinated, through the term $w\,I$.}
\label{fig:SISV-new}
\end{figure}


\qquad We considered a fully-connected population with $N$ individuals. Each agent $i$ in this society carries an opinion $o_{i}$, that is a real number in the range $[-1,+1]$. Positive (negative) values indicate that the position is favorable (unfavorable) to the vaccination campaign. Opinions tending to $+1$ and $-1$ indicate extremist individuals. Finally, opinions near $0$ mean neutral or undecided agents \cite{allan_celia_nuno}. We will consider an epidemic dynamics coupled with the opinion dynamics regarding the vaccination, with the agents being classified follows:  
\begin{itemize}
\item Opinion states: Pro-vaccine (opinion $o_i>0$) or Anti-vaccine (opinion $o_i<0$) individuals;
\item Epidemic compartments: Susceptible (S), Infected (I) or Vaccinated (V) individuals;
\end{itemize}

We define the initial density of positive opinions as $D$, that is a parameter of the model, and in this case the density of negative opinions at the beginning is $1-D$ (for details see Appendix). Figure \ref{fig:SISV-new} shows an esquematic representation of the dynamics. Each Susceptible agent $i$ takes the vaccine with probability $\gamma_{i}$, that is different among the individuals (heterogeneous). This parameter can be viewed as the engagement of the individuals regarding the vaccination campaign, i.e., it measures the tendency of an agent to go to the hospital to take a dose of the vaccine \cite{PiresCrokidakis2017}. In the case a given individual does not take the vaccine, he can become infected with probability $\lambda$ if he make a contact with an Infected individual, as in the standard SIS model. In the same way, an Infected individual becomes Susceptible again with probability $\alpha$. Considering the Vaccinated agents, we considered that the vaccine is not permanent, so a vaccinated agent becomes susceptible again with rate $\phi$, the resusceptibility probability \cite{Zhou_Yicang_Liu2003,Shaw2010}. Summarizing, the individuals can undergo the following transitions among the epidemic compartments:

\begin{itemize}
\item $S \stackrel{\gamma_{i}}{\rightarrow} V$: a Susceptible individual $i$ becomes Vaccinated with probability $\gamma_i$;

\item $S \stackrel{(1-\gamma_{i})\lambda}{\rightarrow} I$:  a Susceptible  individual $i$ becomes Infected with probability $(1-\gamma_i)\lambda$ if he is in contact with an Infected agent;

\item $I \stackrel{\alpha}{\rightarrow} S$: an Infected individual $i$ recovers and becomes susceptible again with probability  $\alpha$;

\item $V \stackrel{\phi}{\rightarrow} S$: a Vaccinated individual $i$ becomes Susceptible again with the resusceptibility probability $\phi$.
\end{itemize}

As previous behavioral change models \cite{CoelhoCodeco2009,Salathe_Bonhoeffer2008,Eames2009,Voinson_Billiard_Alvergne2015,PiresCrokidakis2017,Zuzek2017}, we have not employed a game theory approach, but rather we have considered a mixed  belief-based model that also includes  risk perception of becoming infected (prevalence-based information). This is a plausible hypothesis as was shown in \cite{Voinson_Billiard_Alvergne2015}: \textit{``assumptions of economic rationality and payoff maximization are not mandatory for predicting commonly observed dynamics of vaccination coverage''.} As was also discussed in \cite{Xia2013}, \textit{``if individuals are social followers, the  resulting vaccination coverage would converge to a certain level, depending on individuals' initial level of vaccination willingness rather than the associated costs''.} As empirically observed the individuals are influenced by their social contacts in the process of opinion formation about a vaccination process  \cite{Lau2010,Bish2011}. For those reasons it is important to take opinion dynamics into consideration when studying the spreading of vaccine-preventable diseases.

Discrete opinions can be a good first-order approximation  which sheds light on the problem under investigation and  enables analytical treatment sometimes, but employing continuous opinions is a more realistic approach as mentioned in the introduction.


 We assume, based on kinetic models of collective opinion formation \cite{LCCC2010}, that the opinion dynamics is governed by the equation
 
\begin{equation} \label{eq1}
o_i(t)=o_i(t-1)+\epsilon o_j(t-1)+wI(t-1) ~,
\end{equation}

\noindent
which considers that the  opinion $o_{i}(t)$ of each agent $i$ at an instant $t$ depends on: (i) his previous opinion $ o_i(t-1)$; (ii)  a peer pressure  exerted by a randomly selected agent $j$, modulated by a stochastic variable $\epsilon$ uniformly distributed in the interval $[0,1]$, that introduces heterogeneity in the pairwise interactions; (iii) the proportion of infected agents  $I(t-1)$ modulated by an individuals' risk perception parameter $w$. The opinions are restricted to the range $[ -1,1]$, therefore whenever equation \eqref{eq1} yields $o_{i} > 1 (o_{i}<−1$) it will actually lead the opinion to the extreme $o_{i}=1 (o_{i}=-1$).
The term $wI(t-1)$ can represent a publicly available information released by the mass media (television, radio, newspapers, \ldots ) about the fraction of the population that is infected. It acts as an external field on the opinions, decreasing the vaccine hesitation. Finally, the parameter $w$ takes into account that the risk self-assessment is shaped by several factors, such as personal beliefs, personal experiences and credibility of available information \cite{Cava2005}. The last can be indicated by polls.

We assume that the vaccination probability $\gamma_{i}$  of an agent $i$ is proportional to his opinion about vaccination $o_i$, as follows:

\begin{equation}
\gamma_i(t) = \frac{1+o_{i}(t)}{2} ~.
\end{equation}
\noindent
The above equation ensures that $0\leq\gamma_{i}\leq1$ since $-1\leq o_i\leq1$ and also introduces a high level of heterogeneity in the personal vaccination hesitation $1-\gamma_{i}$ that depends on the corresponding agents' point of view about vaccines.

In the above description of our coupled model we are considering a mixture of prevalence-based and  belief-based model (using the classification proposed in the review \cite{Funk2010}). We stress that our goal is not to model a specific disease spreading,  but rather to investigate the  possible emerging scenarios of a coupled vaccination-opinion dynamics. Note that we assume a mean-field approach (topologically equivalent to a fully-connected network). This means that each agent can interact with any other agent in the population, but only by means of pairwise interactions. Although this is only an approximation, such formulation is not uncommon. It has been discussed that one can capture most of the dynamics of an epidemic, on real social networks, using only mean-field calculations \cite{Bottcher2015}.


\section{Results and Discussion}

Our Monte Carlo simulations are conceived in the framework of an agent-based system, since the individuals (agents) are the primary subject in a social theory \citep{Conte2012}. For this purpose, we have considered populations with $N = 10^4$ agents. As a measure of time we define a Monte Carlo step (mcs) as a visit to each one of the $N$ agents. For sake of simplicity, and without loss of generality, we fixed the recovery probability $\alpha = 0.1$ in all simulations. We use random initial conditions. In order to control the initial condition we used the parameter $D$. This parameter is the initial fraction of pro-vaccine agents, that is $o_i(t=0)>0$ (for details see Appendix). From Eq. \ref{eq1}, it can be noticed that the cases with $D>0.5$ lead to a consensus in $o_i=1 \ \forall i$ since the overall opinion is positive, $\epsilon \geq 0$ and $\omega I(t-1)\geq 0 $. This implies that every agent will certainly vaccinate, which in turn stops the epidemic spreading. For this reason we are more interested in scenarios in which the initial majority is against vaccination, i.e. $D<0.5$.

\begin{figure*}[!ht]
\centering

    \begin{center}
        \begin{subfigure}[t]{0.25\textwidth}
    \includegraphics[width=\textwidth]{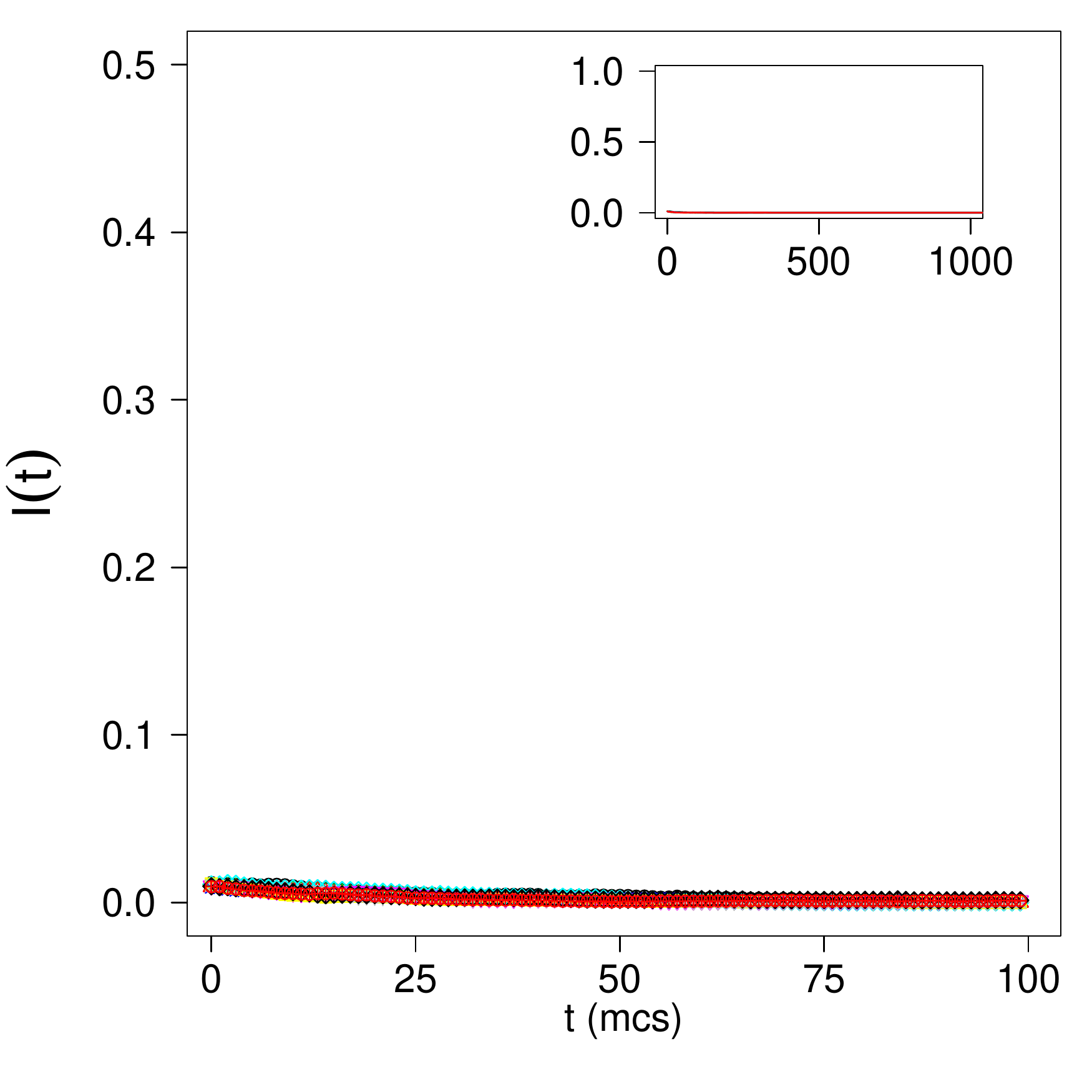}
            \caption{ $\lambda=0.1$ }
        \end{subfigure}
        \begin{subfigure}[t]{0.25\textwidth}
    \includegraphics[width=\textwidth]{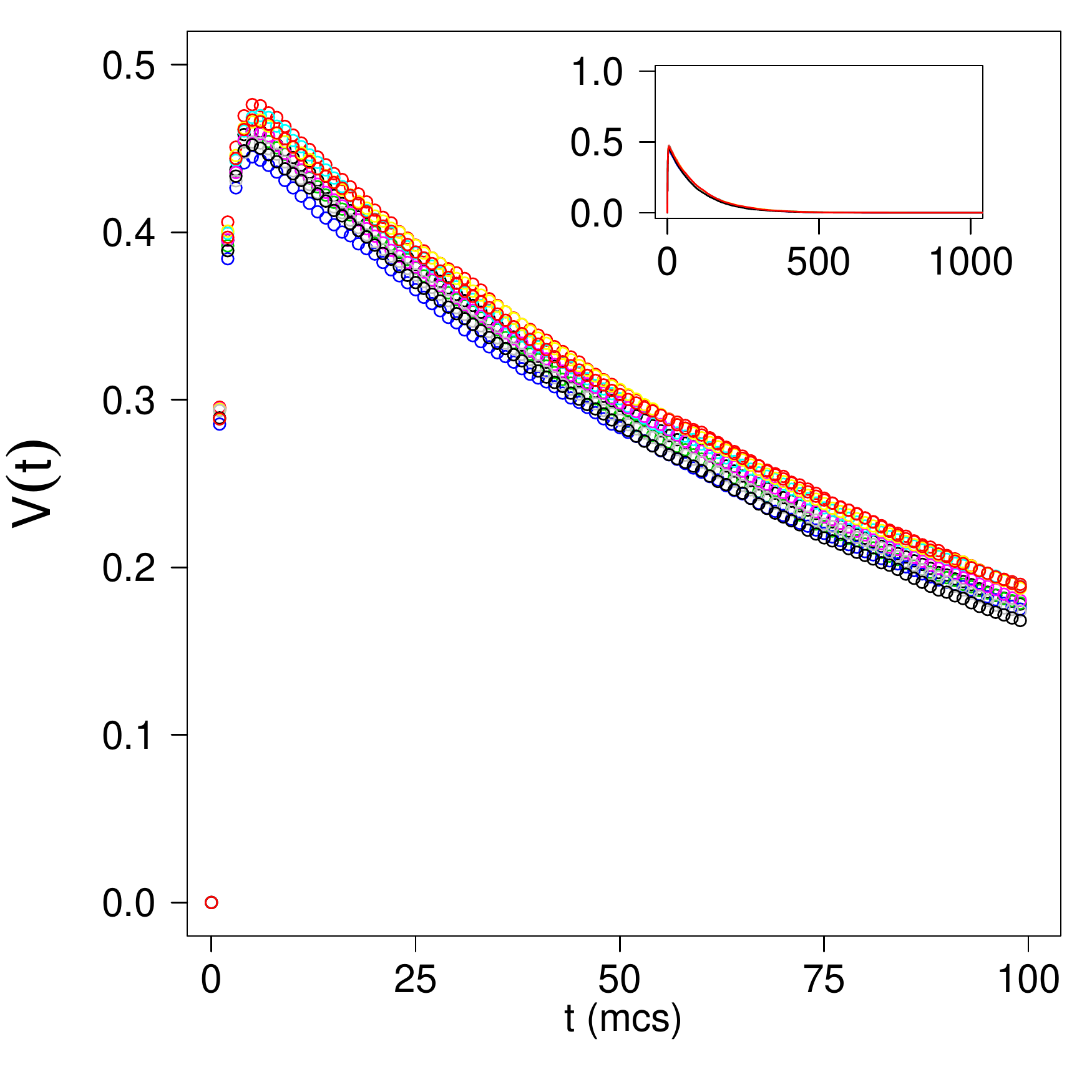}
            \caption{ $\lambda=0.1$ }
        \end{subfigure}        
        \begin{subfigure}[t]{0.25\textwidth}
   \includegraphics[width=\textwidth]{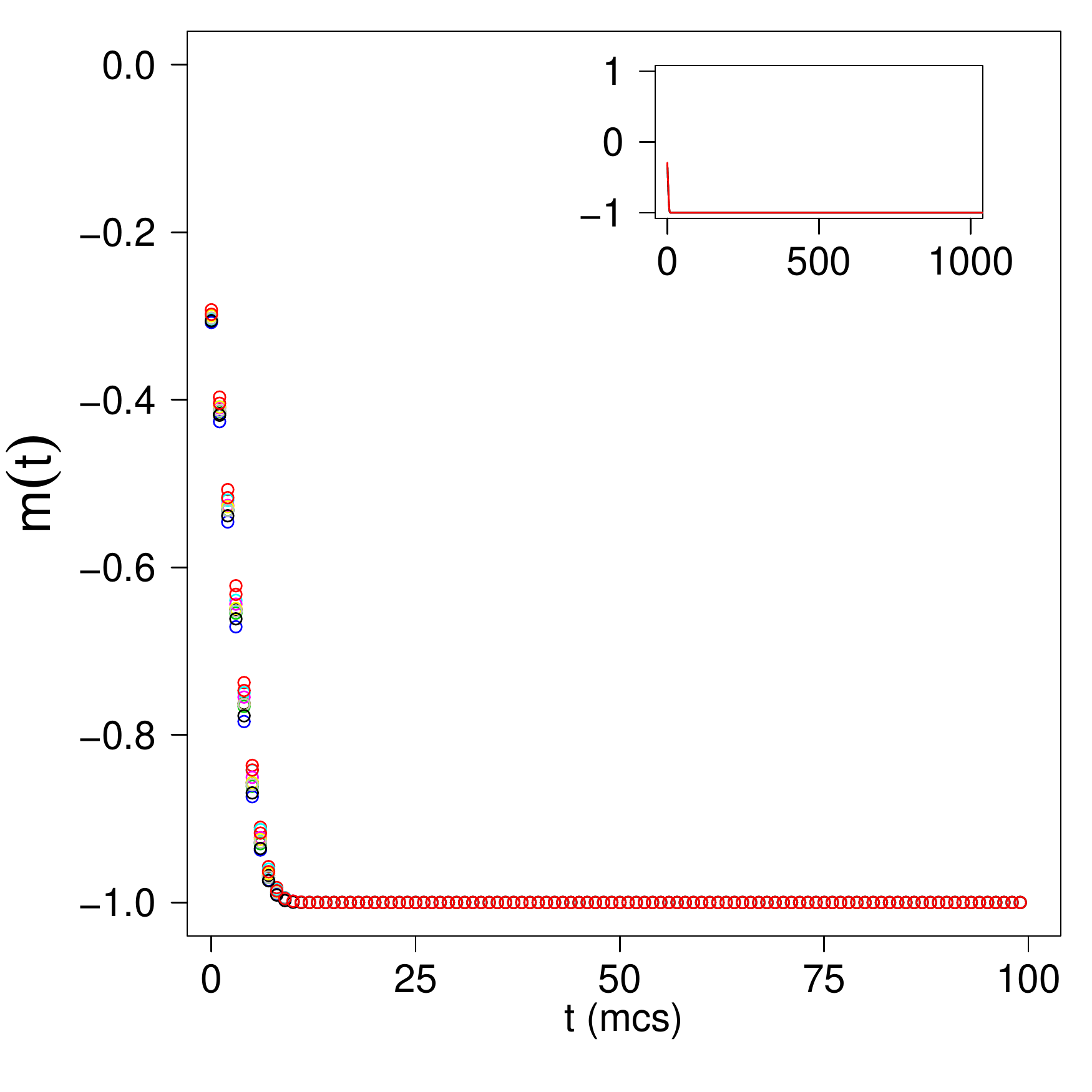}
            \caption{ $\lambda=0.1$ }
        \end{subfigure}

        \begin{subfigure}[t]{0.25\textwidth}
    \includegraphics[width=\textwidth]{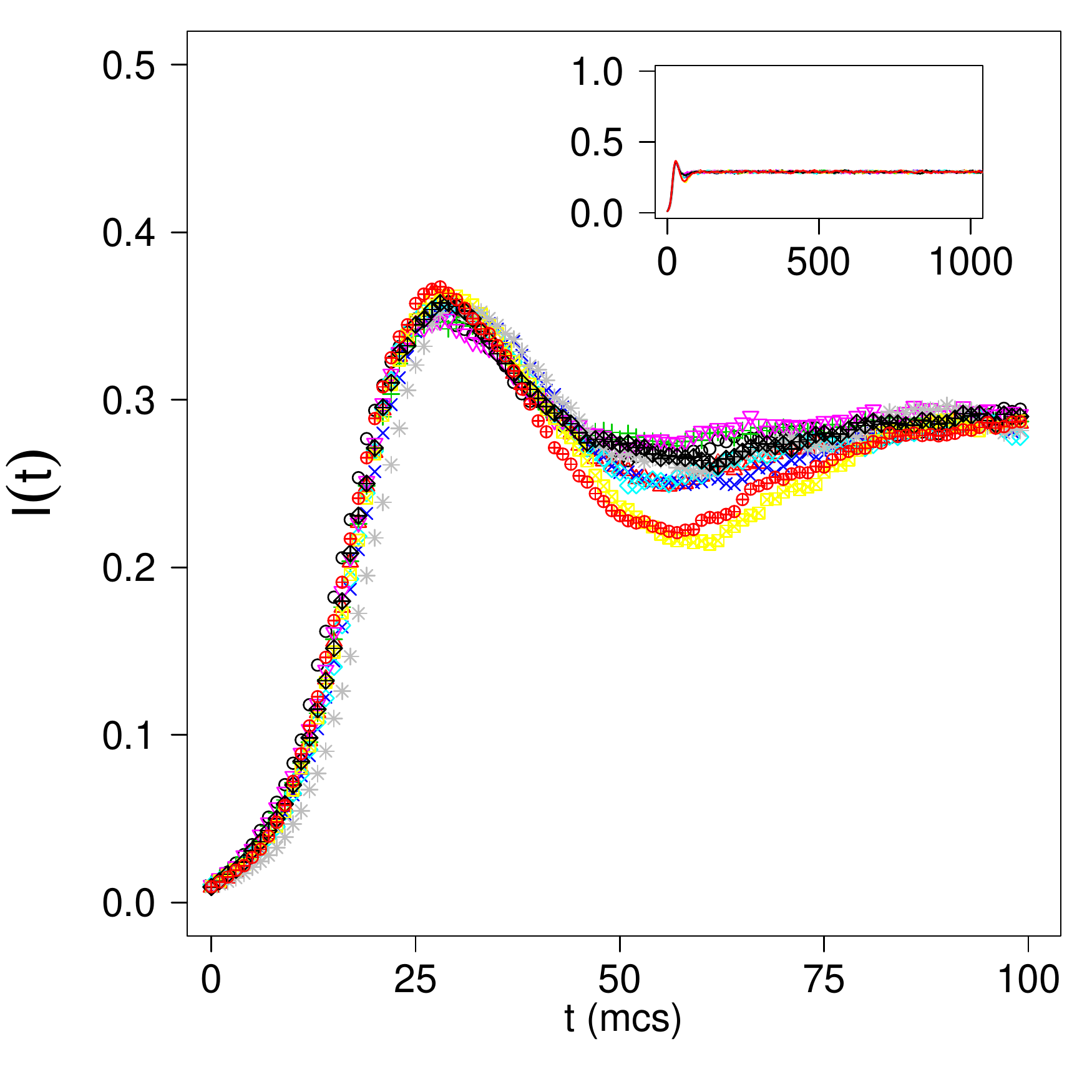}
            \caption{ $\lambda=0.6$ }
        \end{subfigure}
        \begin{subfigure}[t]{0.26\textwidth}
    \includegraphics[width=\textwidth]{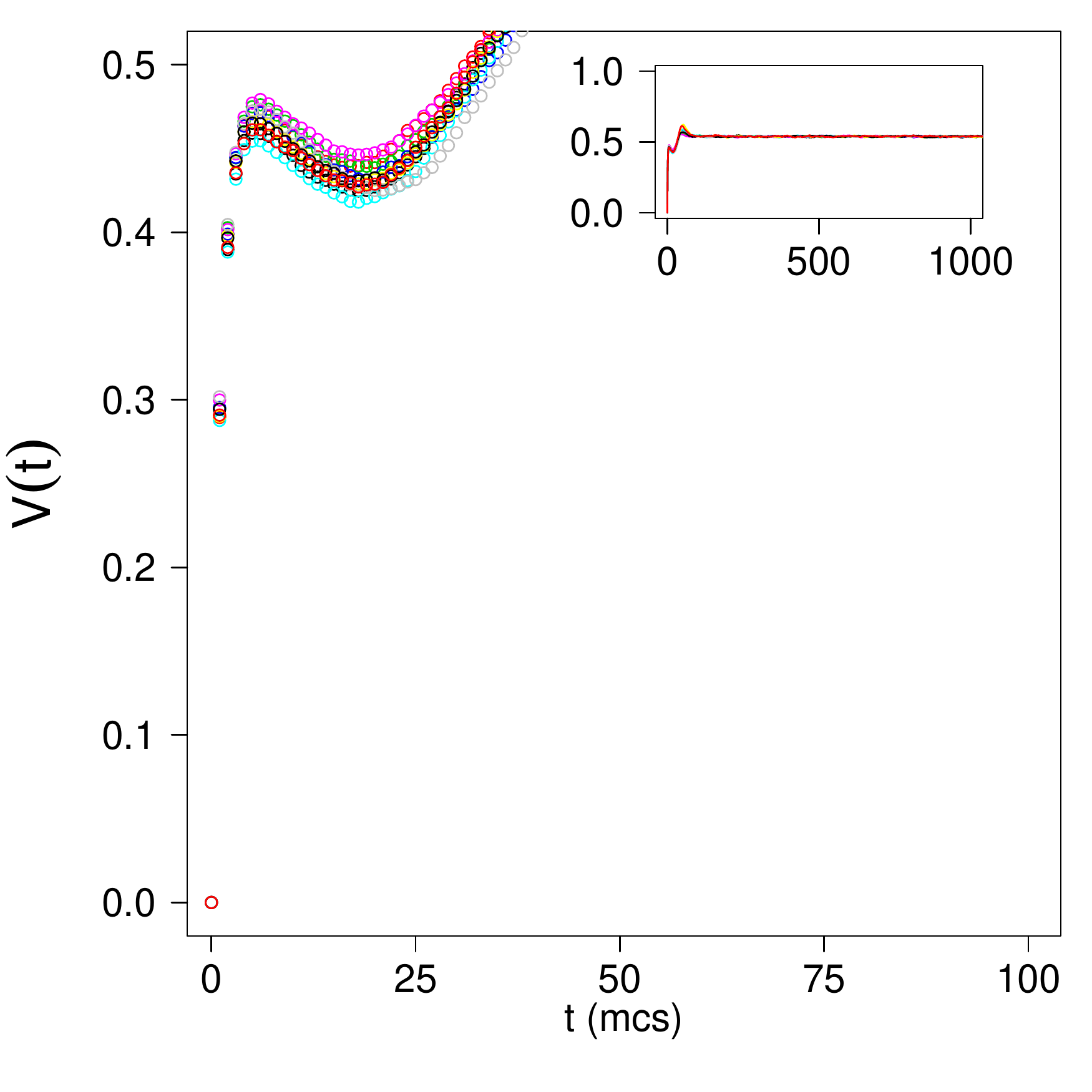}
            \caption{ $\lambda=0.6$ }
        \end{subfigure}        
        \begin{subfigure}[t]{0.25\textwidth}
   \includegraphics[width=\textwidth]{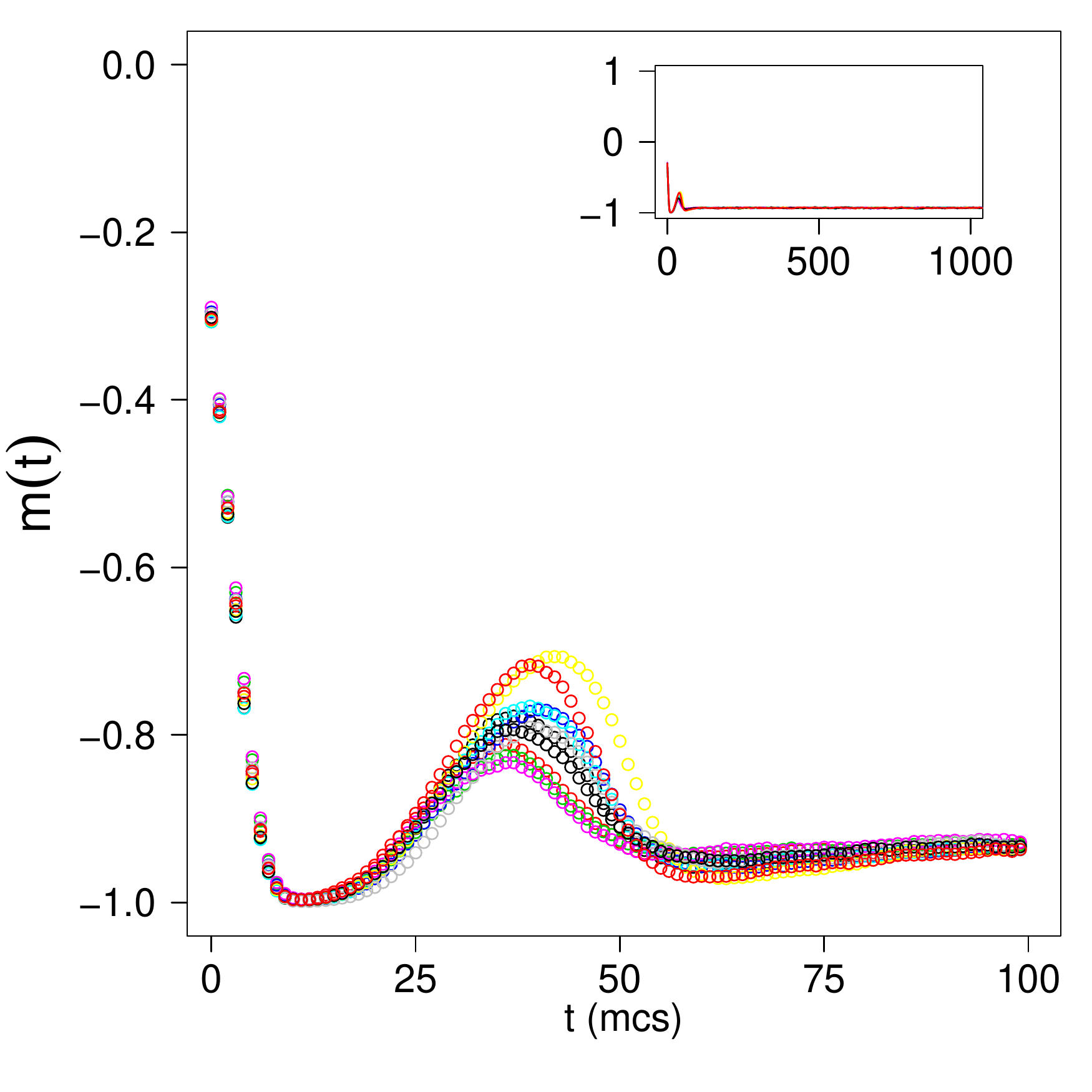}
            \caption{ $\lambda=0.6$ }
        \end{subfigure}

        \begin{subfigure}[t]{0.25\textwidth}
    \includegraphics[width=\textwidth]{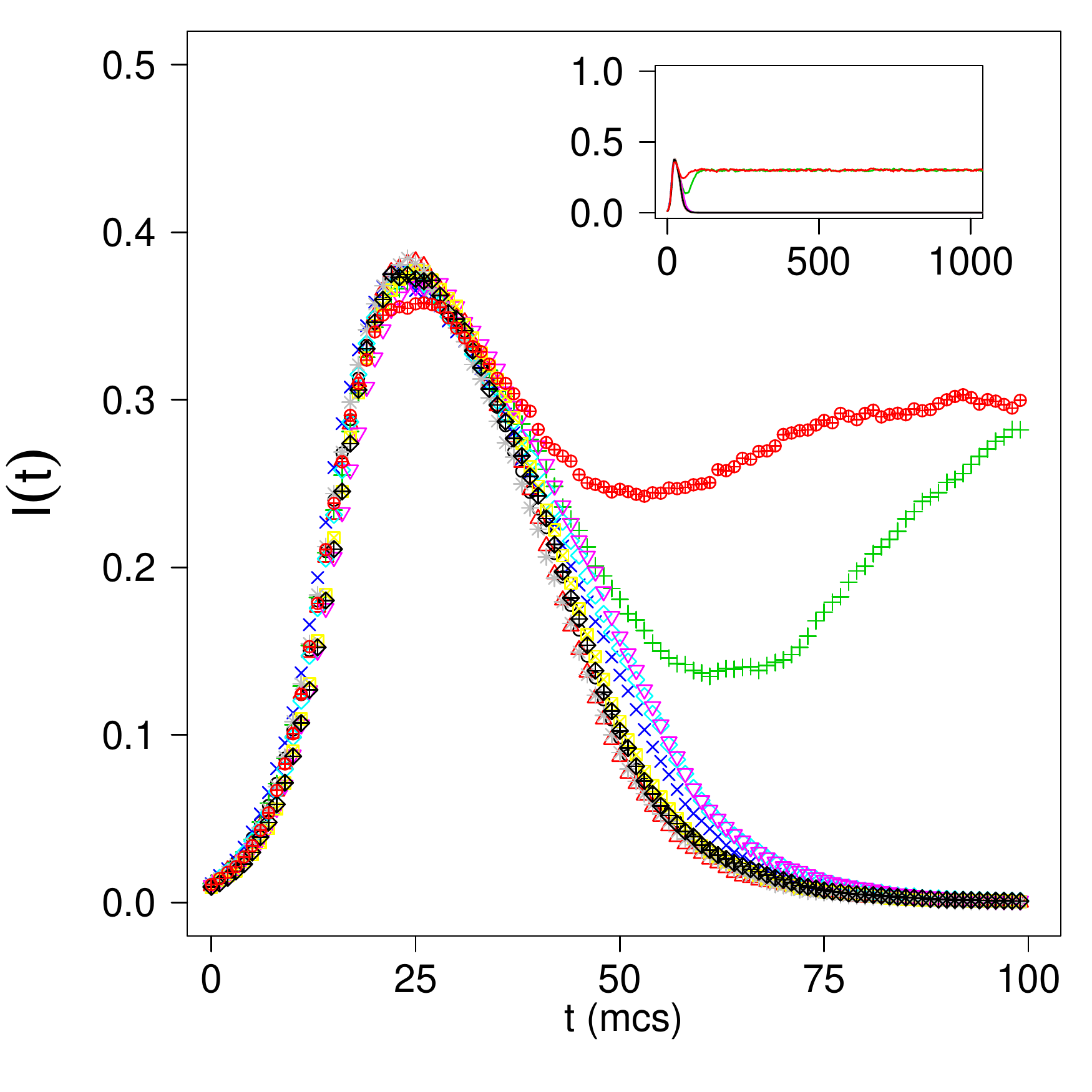}
            \caption{ $\lambda=0.7$ }
        \end{subfigure}
        \begin{subfigure}[t]{0.25\textwidth}
    \includegraphics[width=\textwidth]{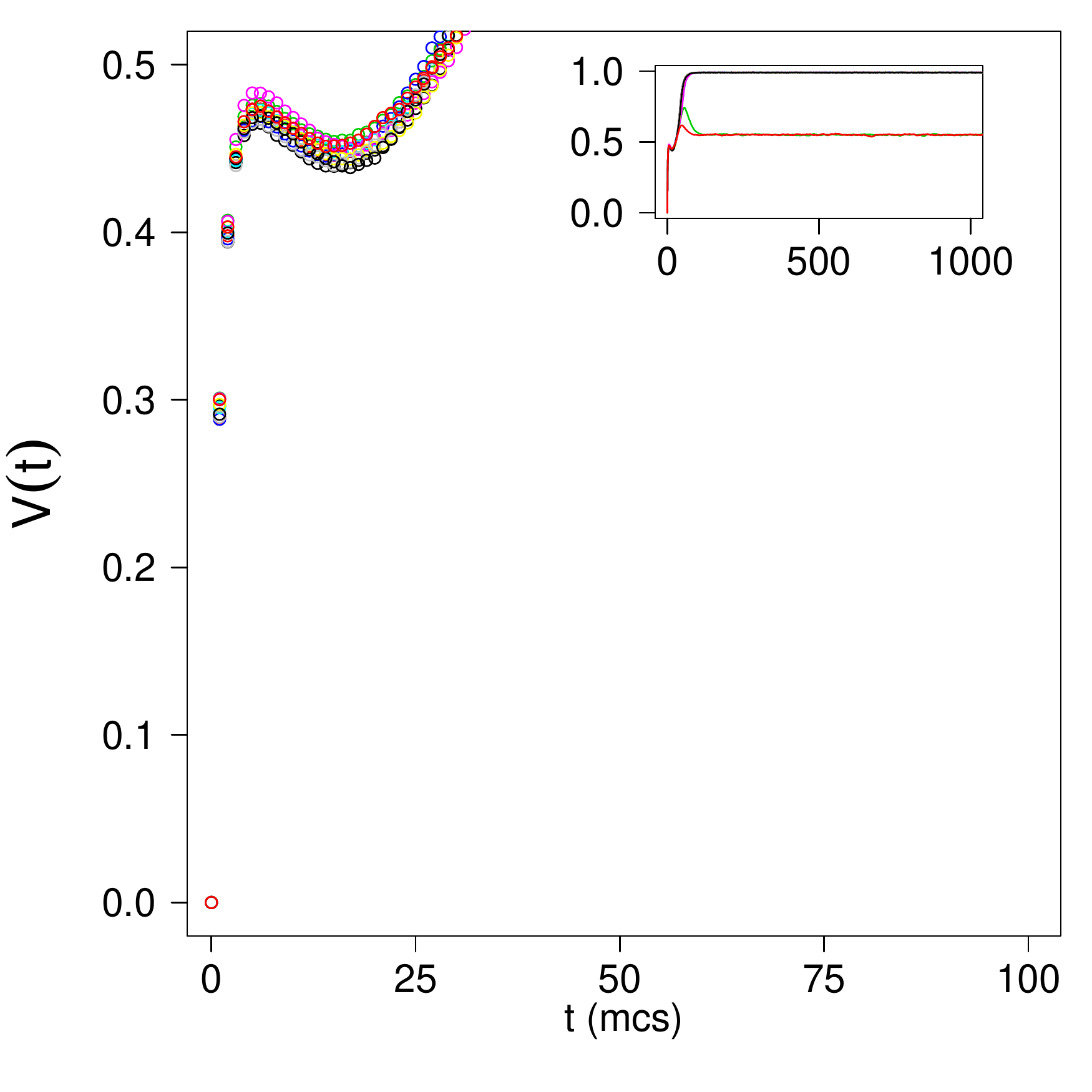}
            \caption{ $\lambda=0.7$ }
        \end{subfigure}        
        \begin{subfigure}[t]{0.25\textwidth}
   \includegraphics[width=\textwidth]{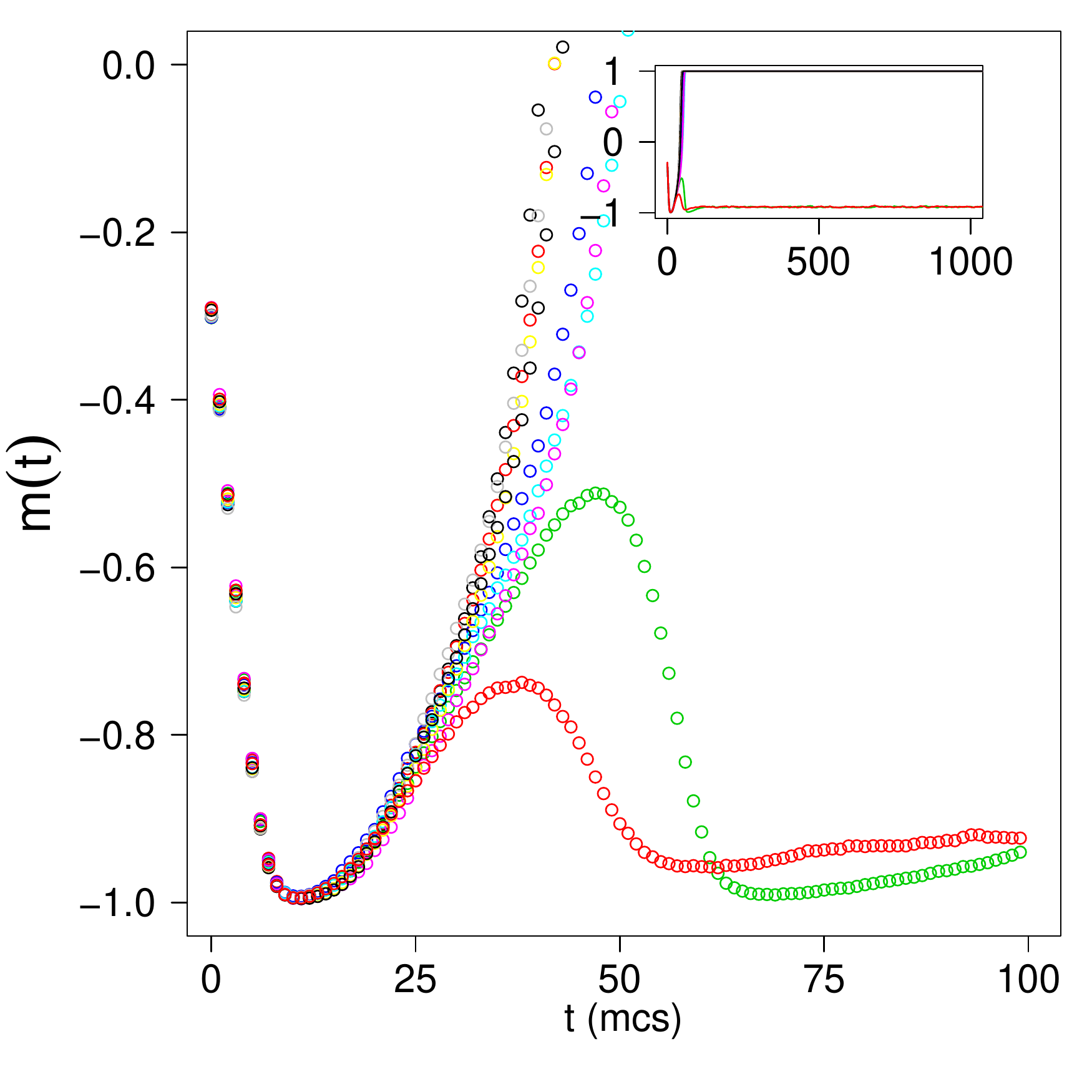}
            \caption{ $\lambda=0.7$ }
        \end{subfigure}

        \begin{subfigure}[t]{0.25\textwidth}
    \includegraphics[width=\textwidth]{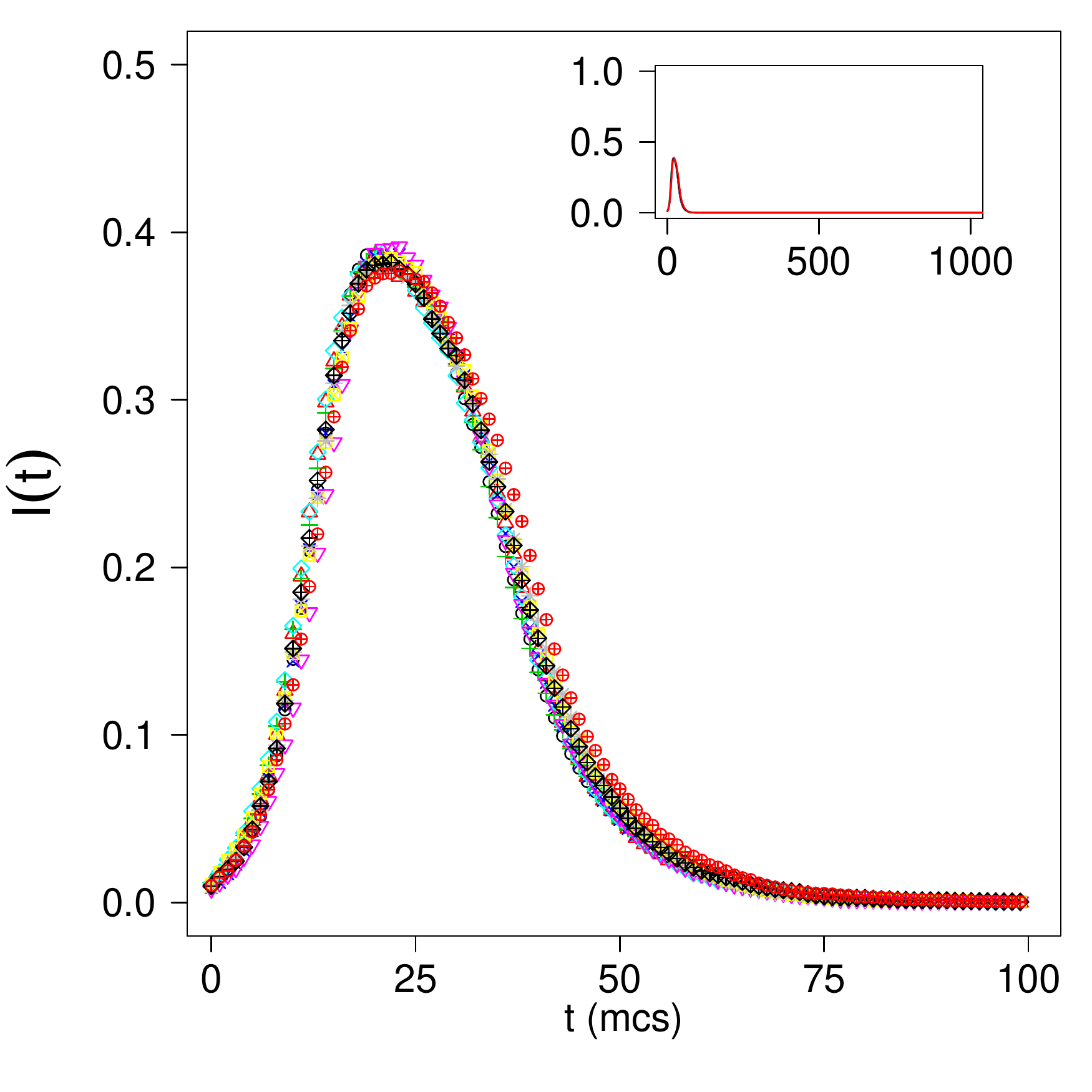}
            \caption{ $\lambda=0.8$ }
        \end{subfigure}
        \begin{subfigure}[t]{0.25\textwidth}
    \includegraphics[width=\textwidth]{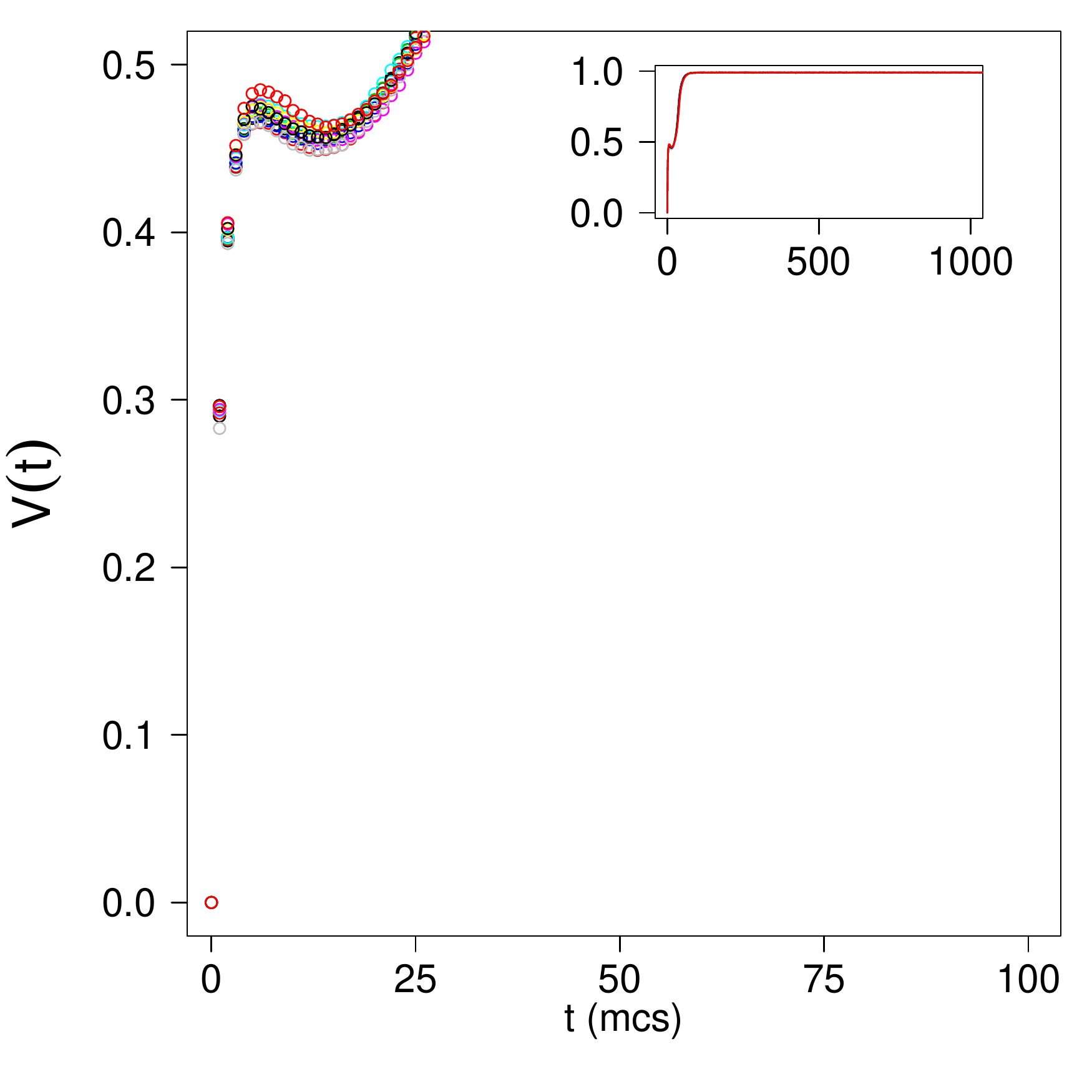}
            \caption{ $\lambda=0.8$ }
        \end{subfigure}        
        \begin{subfigure}[t]{0.25\textwidth}
   \includegraphics[width=\textwidth]{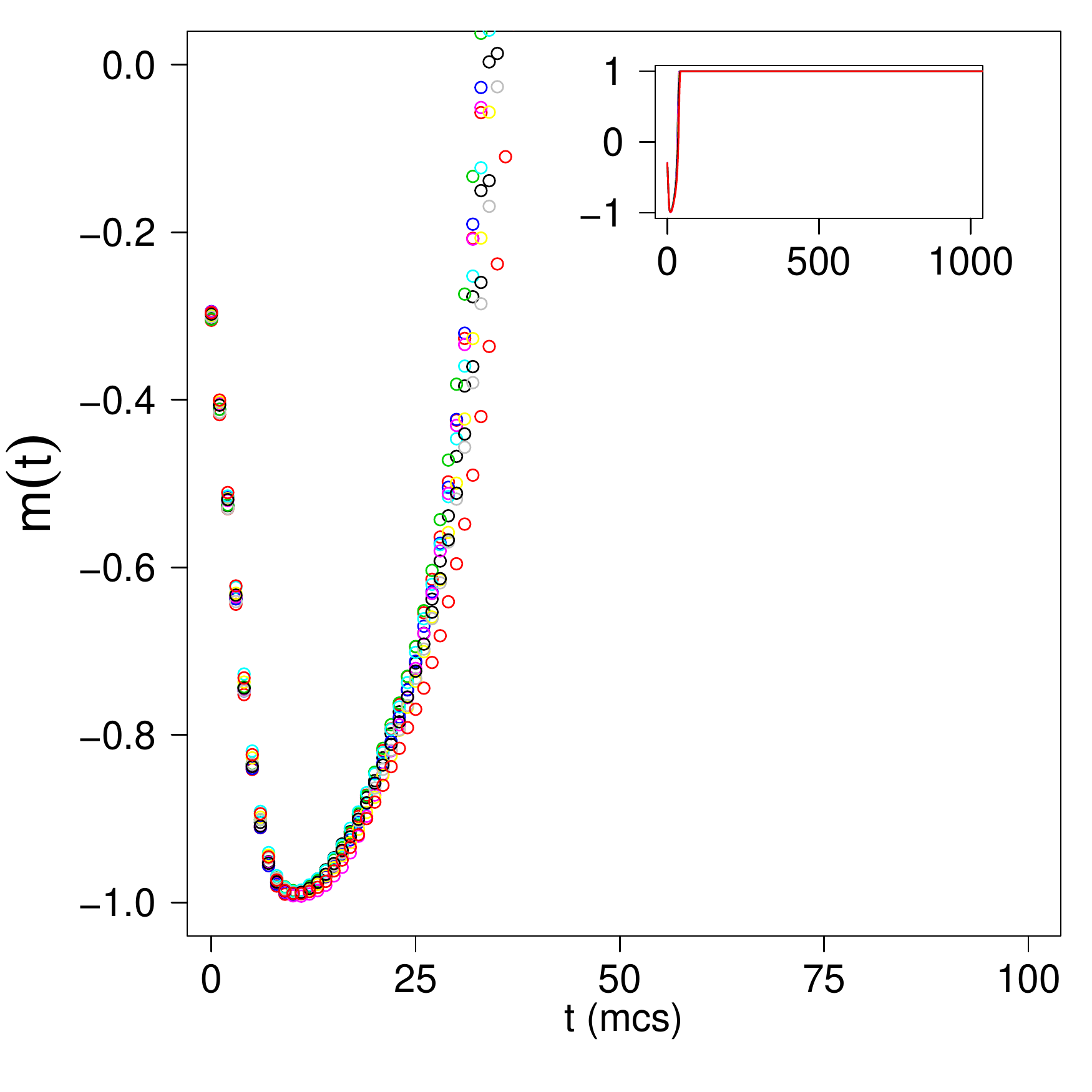}
            \caption{ $\lambda=0.8$ }
        \end{subfigure}

    \end{center}

\caption{ Tuning $\lambda$ leads to different temporal evolutions for $I(t)$, $V(t)$ and $m(t)$. Each time series comes from a single Monte Carlo simulation (realization) without taking any average. Parameters used are $D=0.20$,  $w=0.90$, $\alpha=0.1$,  $\phi=0.01$ and $N=10^{4}$.}  
\label{fig:time-series-1}
\end{figure*}

Lets start by looking at the time series, Fig. \ref{fig:time-series-1}, of the densities of Infected $I$ and Vaccinated $V$ individuals, as well as the average opinion $m$, defined as $m=\sum_{i=1}^{N}o_i/N$. As$\ D = 0.2 $  the initial majority holds a negative view about vaccination. Then in the absence of an external field the social pressure pushes the system towards  a consensus in $o_i=-1 \ \forall i$.  This is exactly what occurs for the infection probability $\lambda=0.1$: the external field $wI$ is turned off when the disease spreading vanishes (let us call it Disease-Free phase I or simply DF I), then all agents end up sharing the same opinion $o_i=-1 \forall i$. This makes the vaccine  coverage vanish after an initial increase. For moderately  epidemic transmissibility, such as $\lambda=0.6$, there is an initial outbreak and a permanent disease spreading in the population (Endemic phase). The consensus $o_i=-1 \ \forall i $ is attained,  but it is not an absorbing state anymore. This happens because of the external field does not vanish ($wI(t)\neq 0$). The agents maintain an intention to vaccinate due to permanent risk perception, and therefore the vaccine coverage does not vanish. In the case of highly contagious diseases, such as $\lambda=0.8$, there is an initial large-scale outbreak. This leads to a strong epidemic pressure that quickly overcomes the social pressure of the initial majority. Therefore, there is a shift in the public opinion causing a further increase in the vaccine coverage. This increase stops the chain of epidemic contagion (henceforth called Disease-Free phase II or just DF II). The temporal evolutions with $\lambda=0.7$ exhibit two distinct stable steady states (bistable solutions). 
The aforementioned feedback between the social and epidemic pressures is responsible for the emergence of the bistability in Fig. 2. Also note that the randomness in the initial condition and in the dynamics are very important to reveal this bistability. 
   
\begin{figure*}[!ht]
\centering

    \begin{center}
        \begin{subfigure}[t]{0.31\textwidth}
            \includegraphics[width=\textwidth]{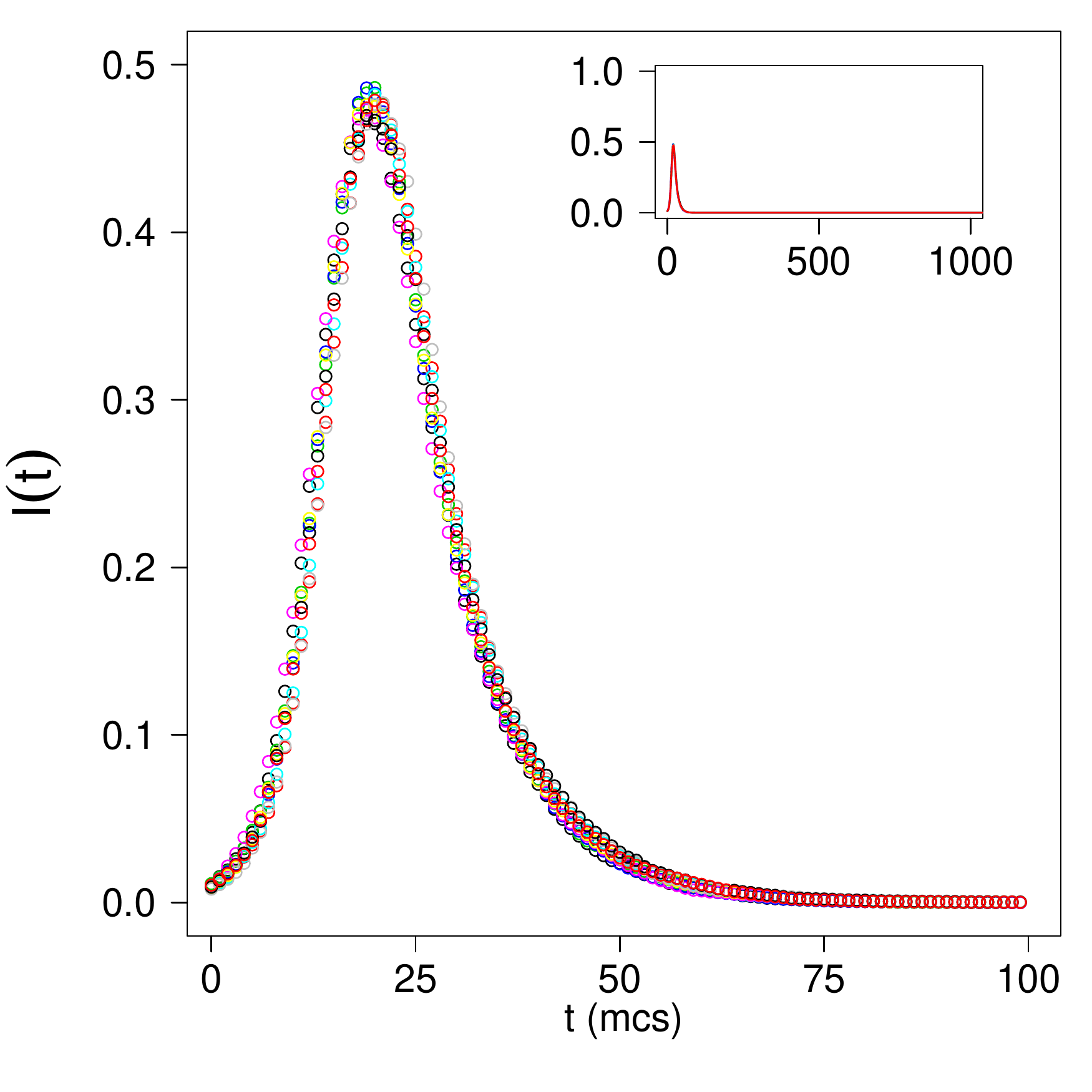}
            \caption{$D=0\%$}
        \end{subfigure}
        \begin{subfigure}[t]{0.31\textwidth}
            \includegraphics[width=\textwidth]{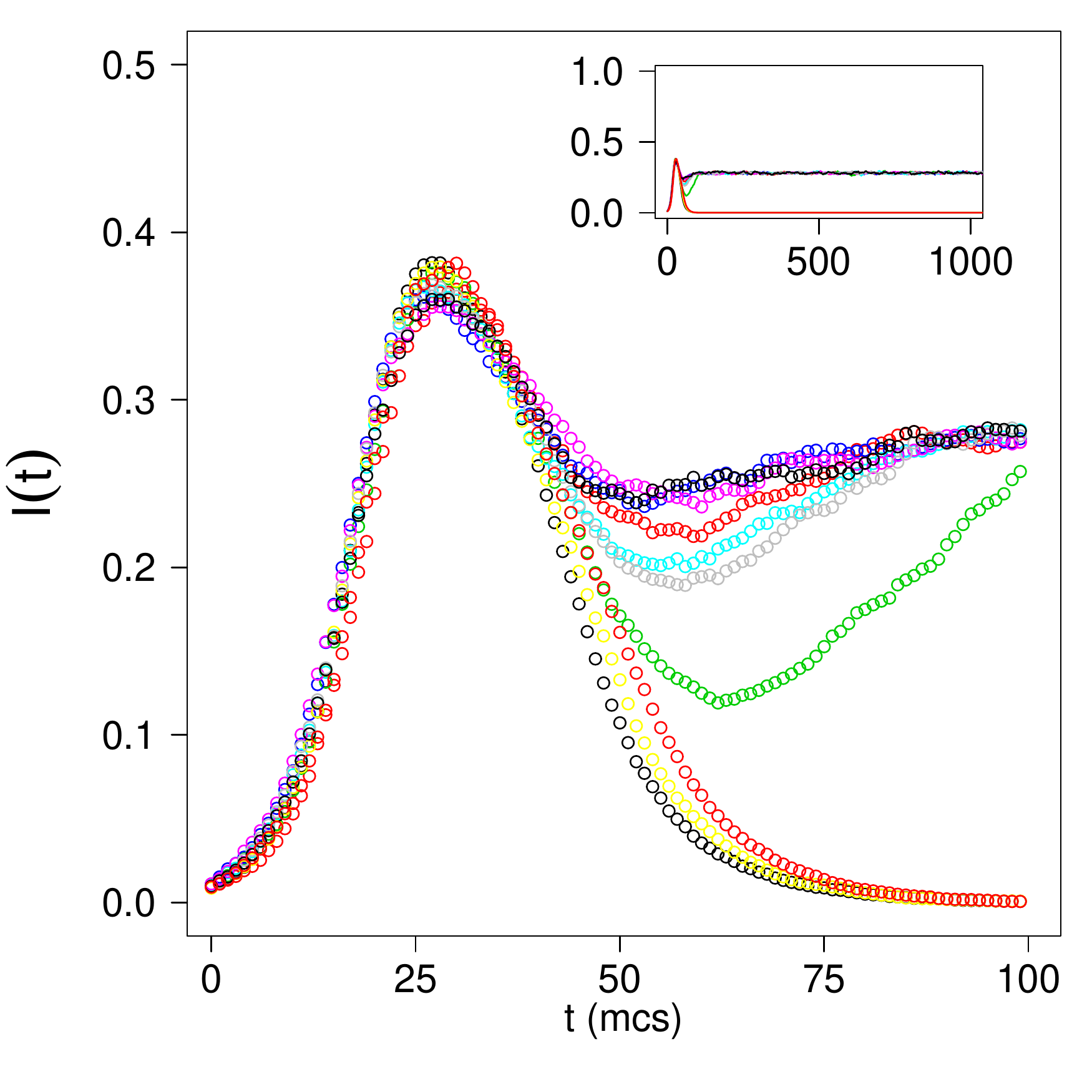}
            \caption{$D=16\%$}
        \end{subfigure}
        \begin{subfigure}[t]{0.31\textwidth}
            \includegraphics[width=\textwidth]{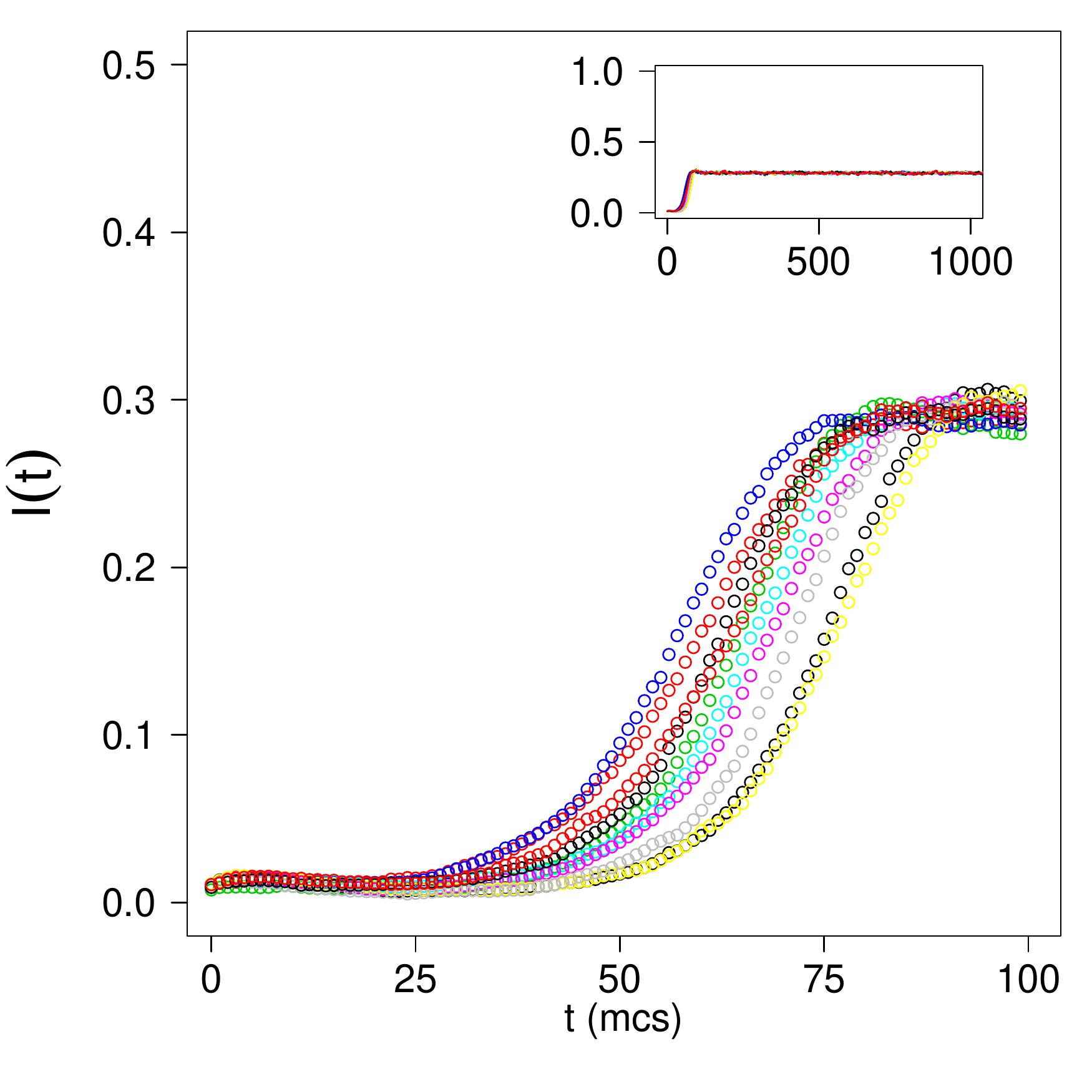}
            \caption{$D=45\%$}
        \end{subfigure}
        
        \begin{subfigure}[t]{0.31\textwidth}
            \includegraphics[width=\textwidth]{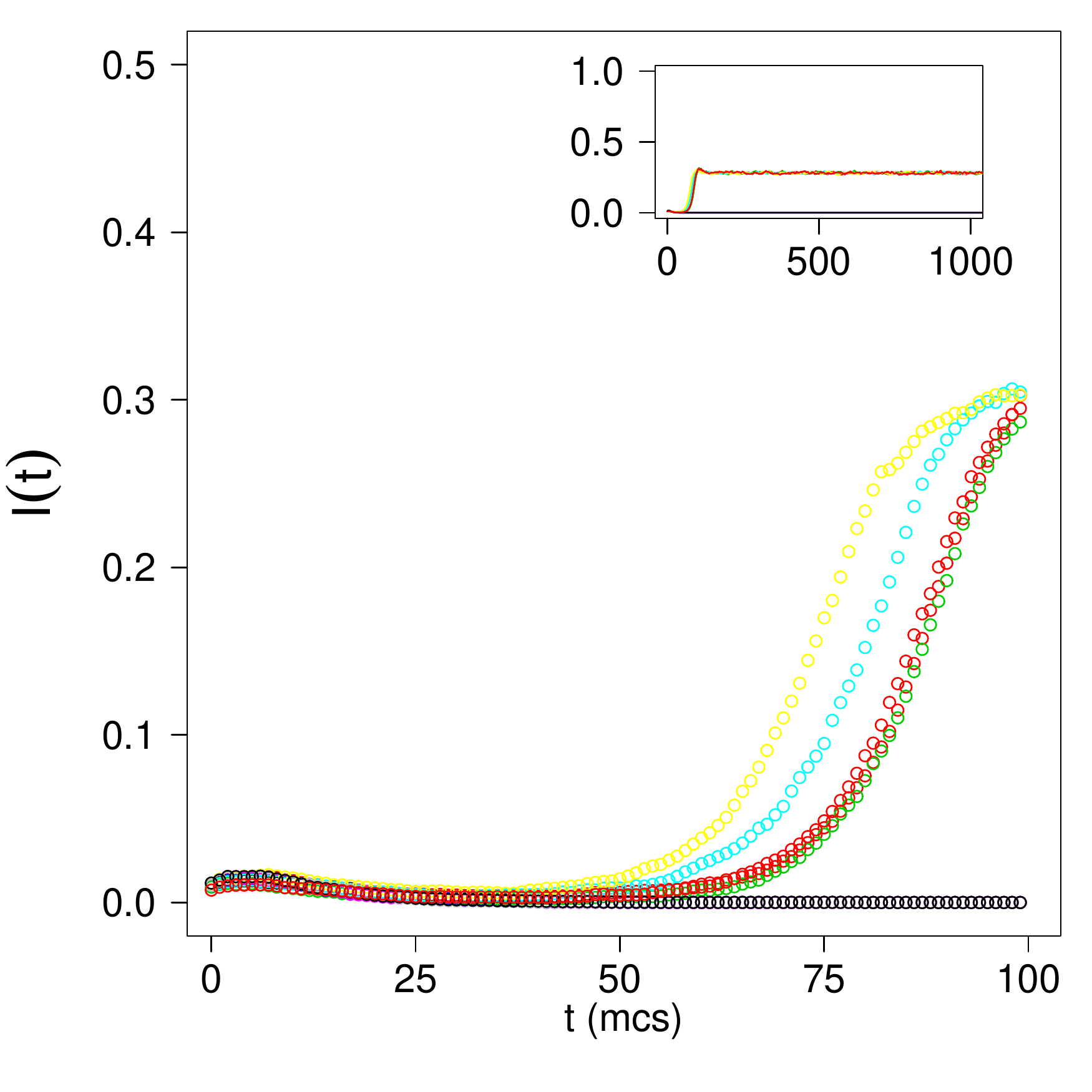}
            \caption{$D=47\%$}
        \end{subfigure}
        \begin{subfigure}[t]{0.31\textwidth}
            \includegraphics[width=\textwidth]{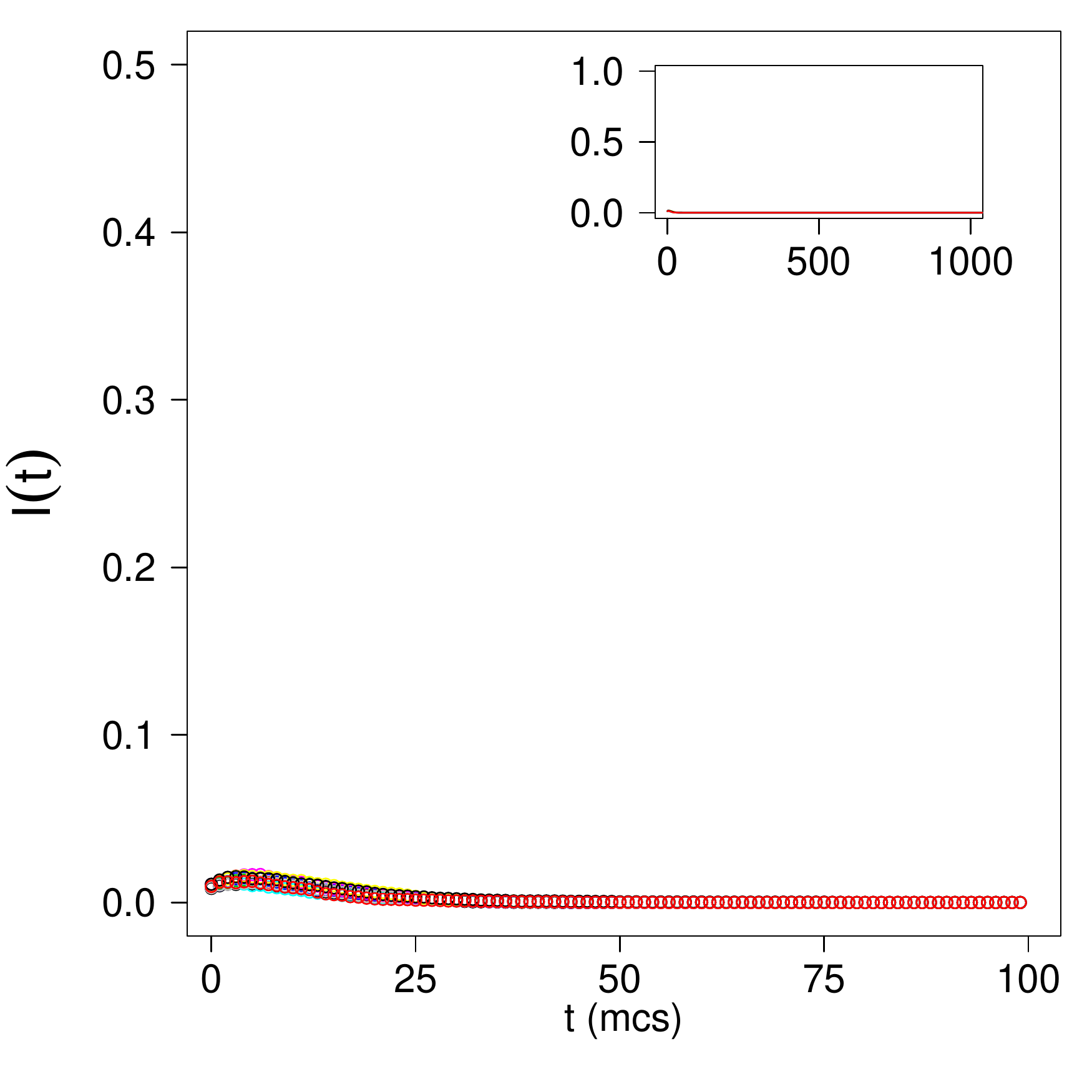}
            \caption{$D=49\%$}
        \end{subfigure}
    \end{center}

\caption{
Time series of $I(t)$ for several values of the initial density $D$ of positive opinions.  For clarity we show only $10$ time series in each subplot. Each time series comes from a single Monte Carlo simulation (realization) without taking any average. Parameters used are $\lambda=0.55$,  $w=0.90$, $\alpha=0.1$,  $\phi=0.01$ and $N=10^{4}$.   }
\label{fig:time-series-2}
\end{figure*}

Let us now turn our attention to Fig. \ref{fig:time-series-2}. Note that $I(t)$ depends on the initial condition (parameter $D$) of the opinion dynamics. This finding is in agreement with previous studies that have used discrete opinion models \cite{PiresCrokidakis2017,Xia2013,Wu2013}. Very interestingly, Fig. \ref{fig:time-series-2} shows that an increment in the initial density of pro-vaccine agents, from $D=0\%$ to $D=45\%$, has a twofold effect: it can lead to a smaller epidemic outbreak in the short term (positive effect). But, on the other hand it can also favor the long-term prevalence of the disease (negative effect). The underlying mechanism behind this result is the competition of social and epidemic pressures as explained for the Fig. \ref{fig:time-series-1} with the exception that now the tuning parameter is $D$ instead of $\lambda$.  The counterintuitive  result showed in Fig. \ref{fig:time-series-2} agrees with a previous counterintuitive finding \cite{Zhang2013} showing that, in determined scenarios, a better condition may lead to a worse outcome for the population. 

The emergence of outcomes that are positive and negative is a consequence of the interplay between the disease spreading and the  changeable human behavior. Indeed, this kind of ``double-edged sword'' effect has been considered a universal feature of coupled behavior-disease models, and has been reported in many distinct contexts as reviewed in \cite{Wang2016}.

 From a policy-oriented perspective, interventions designed to improve the initial public opinion about vaccination (increase $D$) should be implemented very cautiously since such interventions can achieve short-term goals (such as prevention of a large-scale epidemic outbreak), but it can hamper long-term targets (for instance disease eradication). This result is in line with the conclusion in \cite{Wang2015} that \textit{``any disease-control policy should be exercised with extreme care: its success depends on the complex interplay among the intrinsic mathematical rules of epidemic spreading, governmental policies, and behavioral responses of individuals''}. Moreover, there are other interventions such as social distancing that can produce negative consequences \cite{Maharaj_Kleczkowski2012}. There are also control strategies that are only efficient under infeasible scenarios, such as self-isolation as exemplified in \cite{CrokidakisQueiros2012}. Let us mention that  due to the feedback between epidemics and human behaviors even the application of multiple control strategies can yield unexpected outcomes for the society \cite{Zhang2013}. Counterintuitive results can also arise when subsidy policies are implemented in vaccination campaigns as shown in \cite{Zhang2017} where the authors found that there are scenarios in which the final epidemic size increases with the proportion of subsidized individuals. 
  
\begin{figure*}[!ht]
\centering

    \begin{center}
    
            \begin{subfigure}[t]{0.30\textwidth}
        \includegraphics[width=\textwidth]{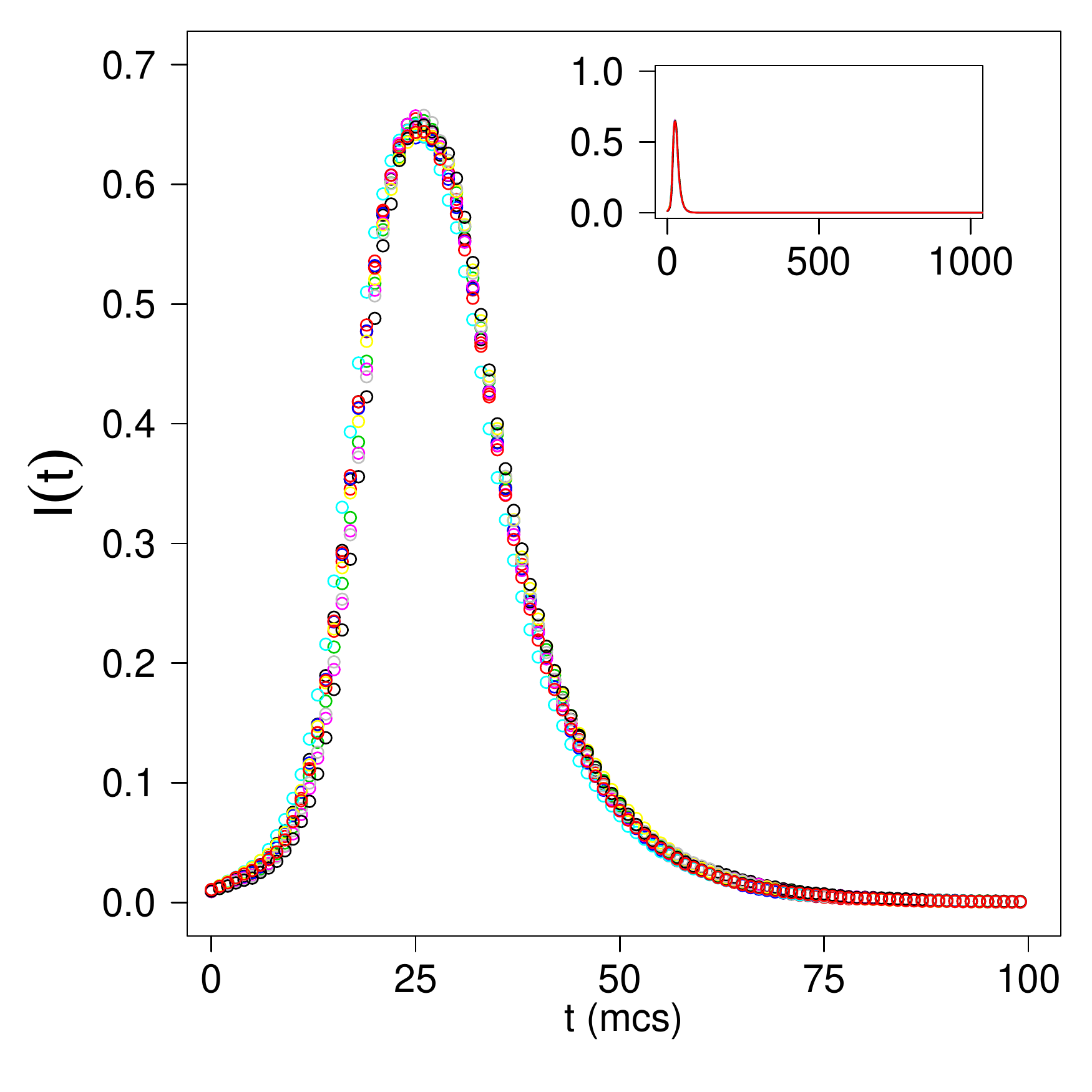}
                \caption{ $\phi=0.1$ }
            \end{subfigure}
            \begin{subfigure}[t]{0.30\textwidth}
        \includegraphics[width=\textwidth]{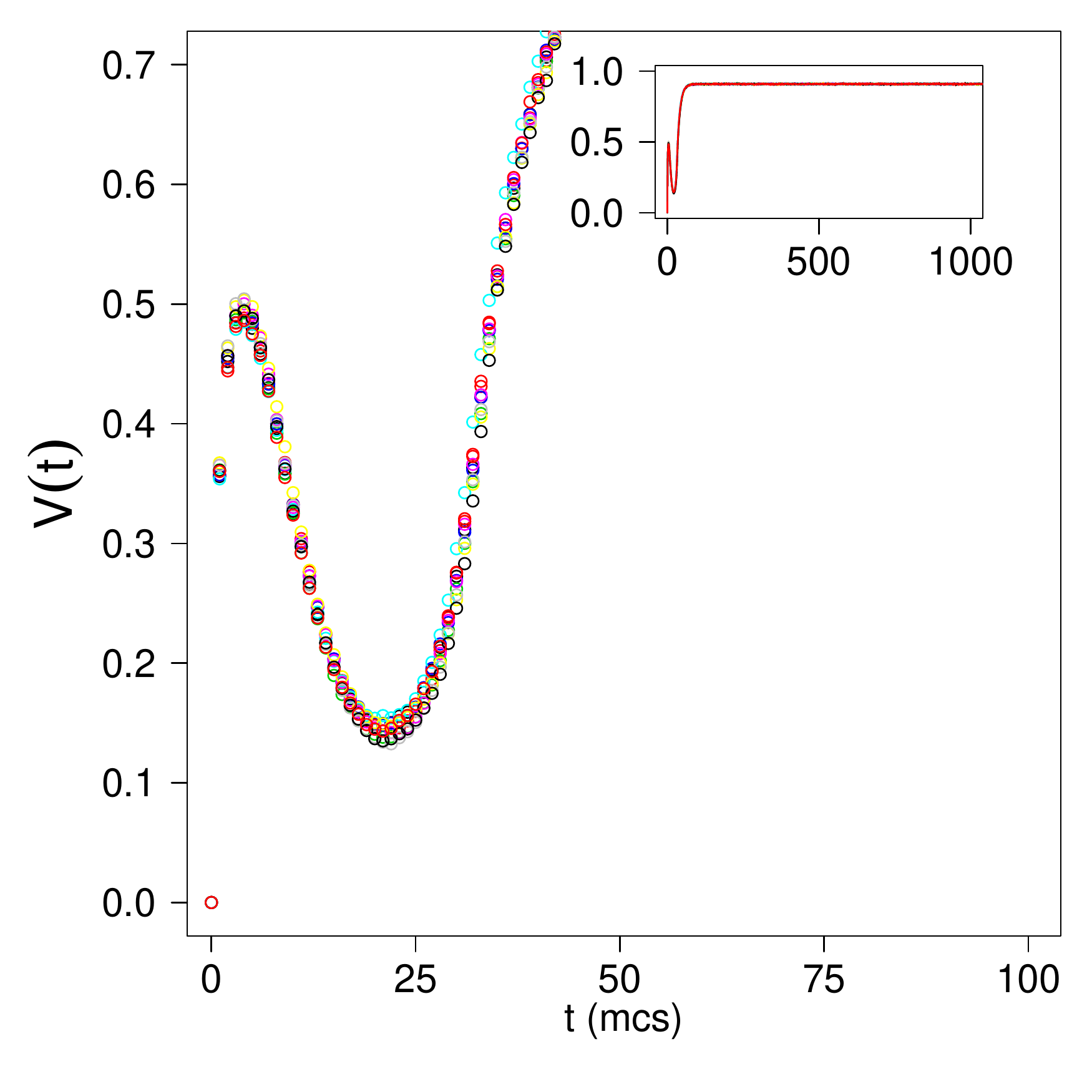}
                \caption{ $\phi=0.1$ }
            \end{subfigure}        
            \begin{subfigure}[t]{0.30\textwidth}
       \includegraphics[width=\textwidth]{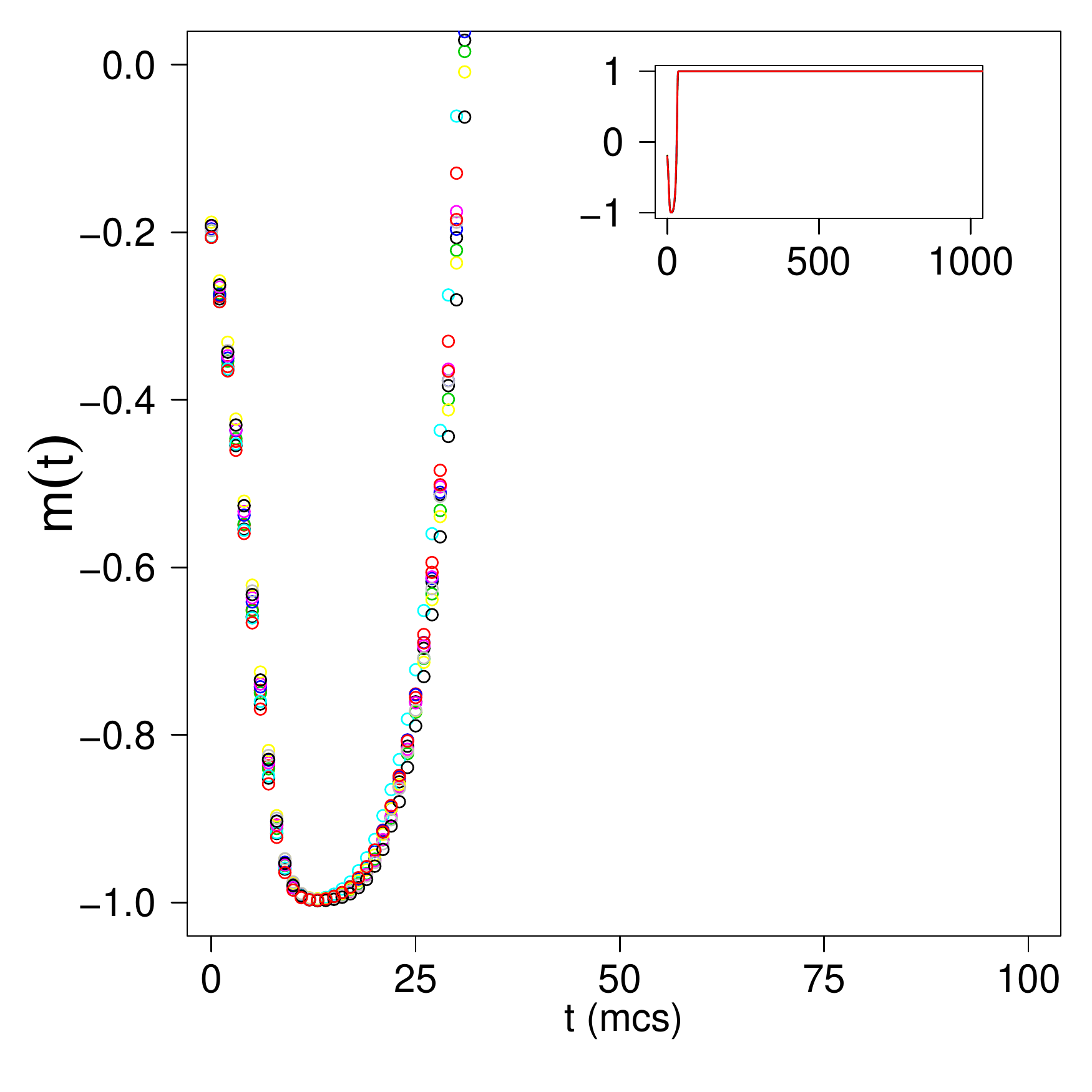}
                \caption{ $\phi=0.1$ }
            \end{subfigure}

        \begin{subfigure}[t]{0.30\textwidth}
    \includegraphics[width=\textwidth]{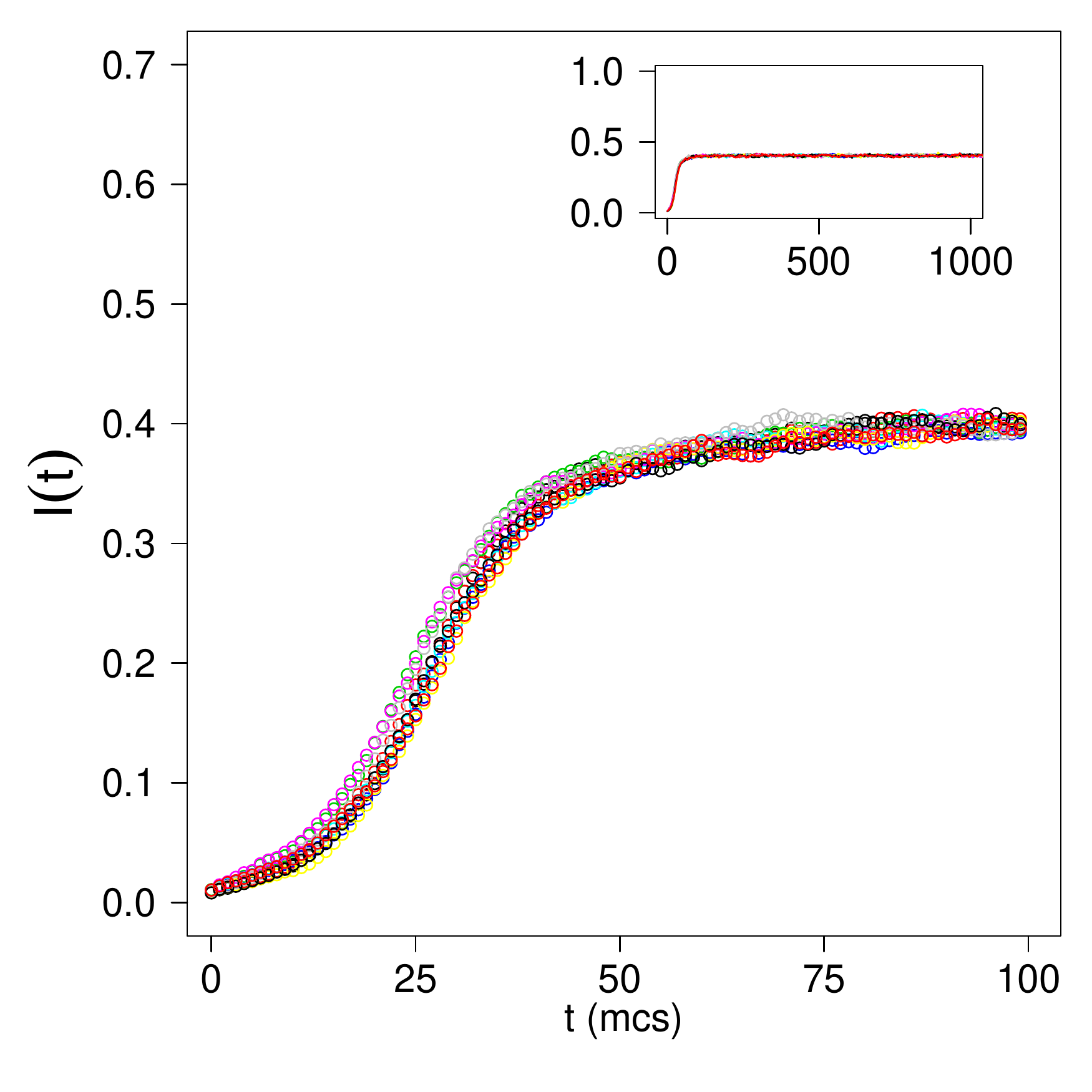}
            \caption{ $\phi=0.01$ }
        \end{subfigure}
        \begin{subfigure}[t]{0.30\textwidth}
    \includegraphics[width=\textwidth]{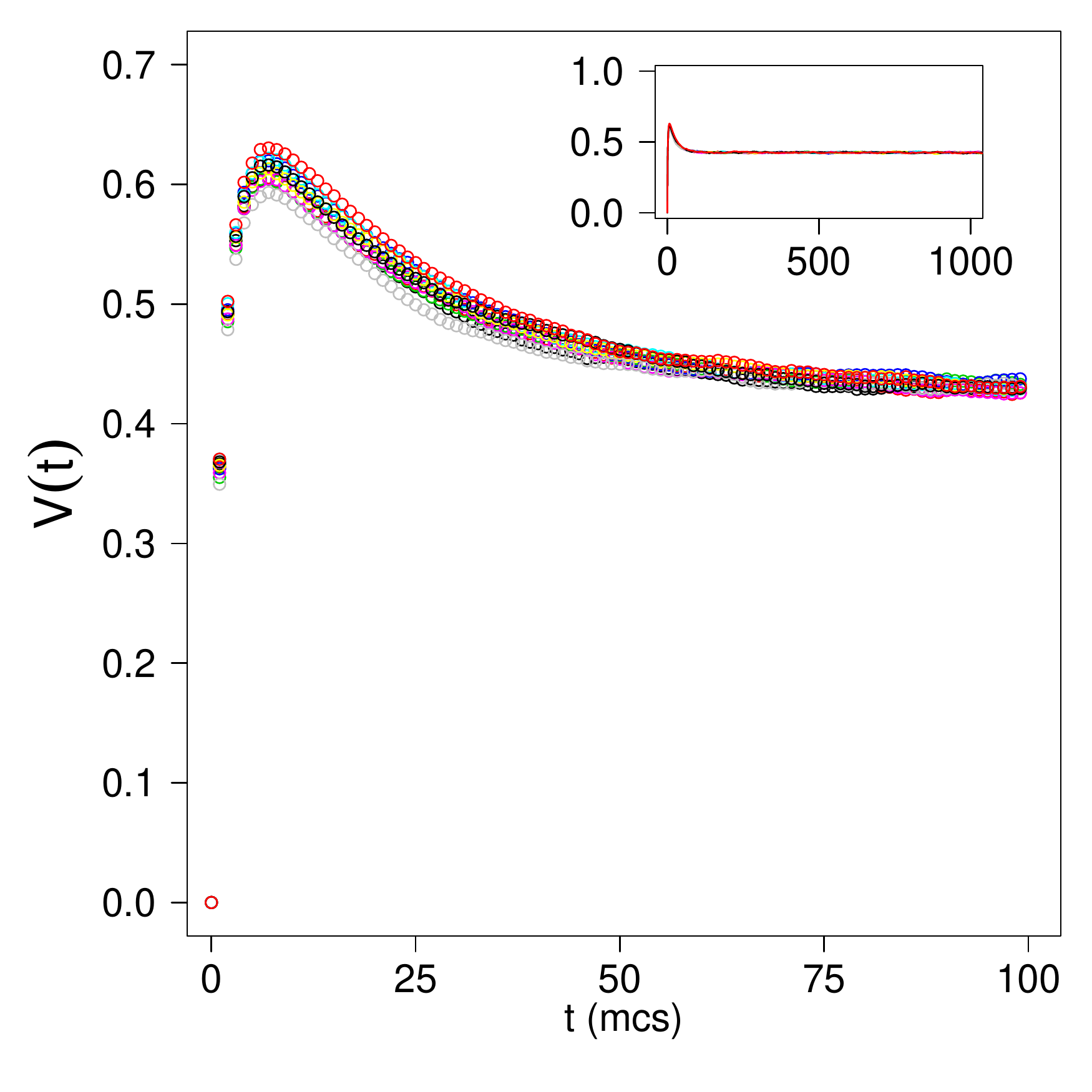}
            \caption{ $\phi=0.01$ }
        \end{subfigure}        
        \begin{subfigure}[t]{0.30\textwidth}
   \includegraphics[width=\textwidth]{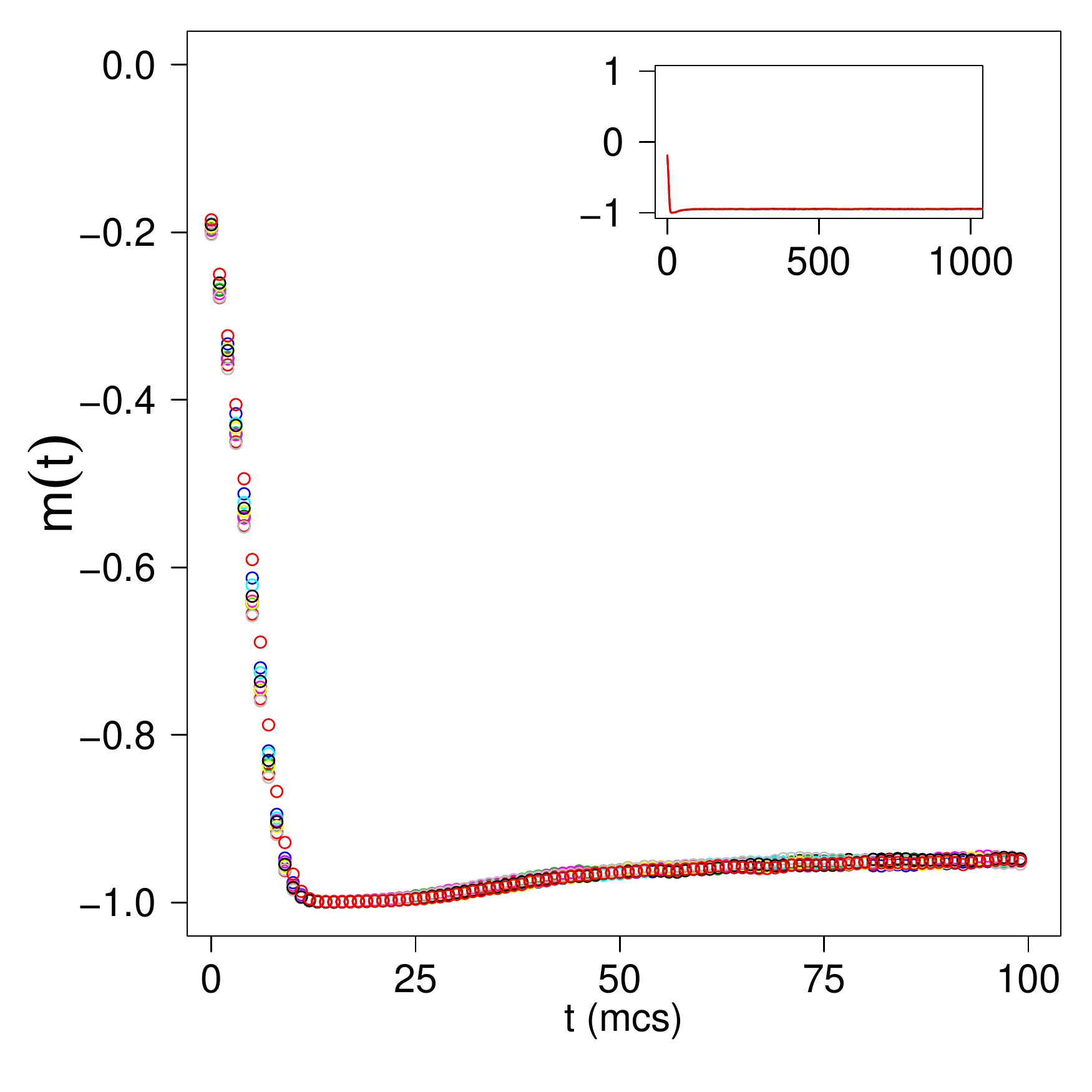}
            \caption{ $\phi=0.01$ }
        \end{subfigure}
    \end{center}

\caption{ Tuning $\phi$ leads to different temporal evolutions for $I(t)$, $V(t)$ and $m(t)$.  For clarity we show only $10$ time series in each subplot. Each time series comes from a single Monte Carlo simulation (realization) without taking any average. Parameters used are $D=0.3$, $\lambda=0.60$,  $w=0.60$, $\alpha=0.1$ and $N=10^{4}$.} 
\label{fig:time-series-3}
\end{figure*}

Fig. \ref{fig:time-series-3} disentangles the role played by the vaccine efficiency ($\phi$). During the first stage of the epidemic spreading there is an increment in the vaccine coverage, due to the initial vaccine intention. The scenarios with $\phi=0.01$  present a more pronounced peak in $V(t)$ than the corresponding scenario with $\phi=0.1$. Higher levels of vaccine coverage induce a stronger slowdown of the epidemic spreading. This in turn weakens the external field ($wI$), and hence reduces the overall vaccination intention. Therefore,  more effective vaccines (smaller $\phi$) can decrease the long-term vaccine coverage. This seems counterintuitive at first, but it is in agreement with empirical observations. As observed in \cite{LeeMale2011,Larson2011}, effective vaccines can become a victim of their own success.

\begin{figure*}[!ht]
\centering

    \begin{center}    
      \begin{subfigure}[t]{0.30\textwidth}
        \includegraphics[width=\textwidth]{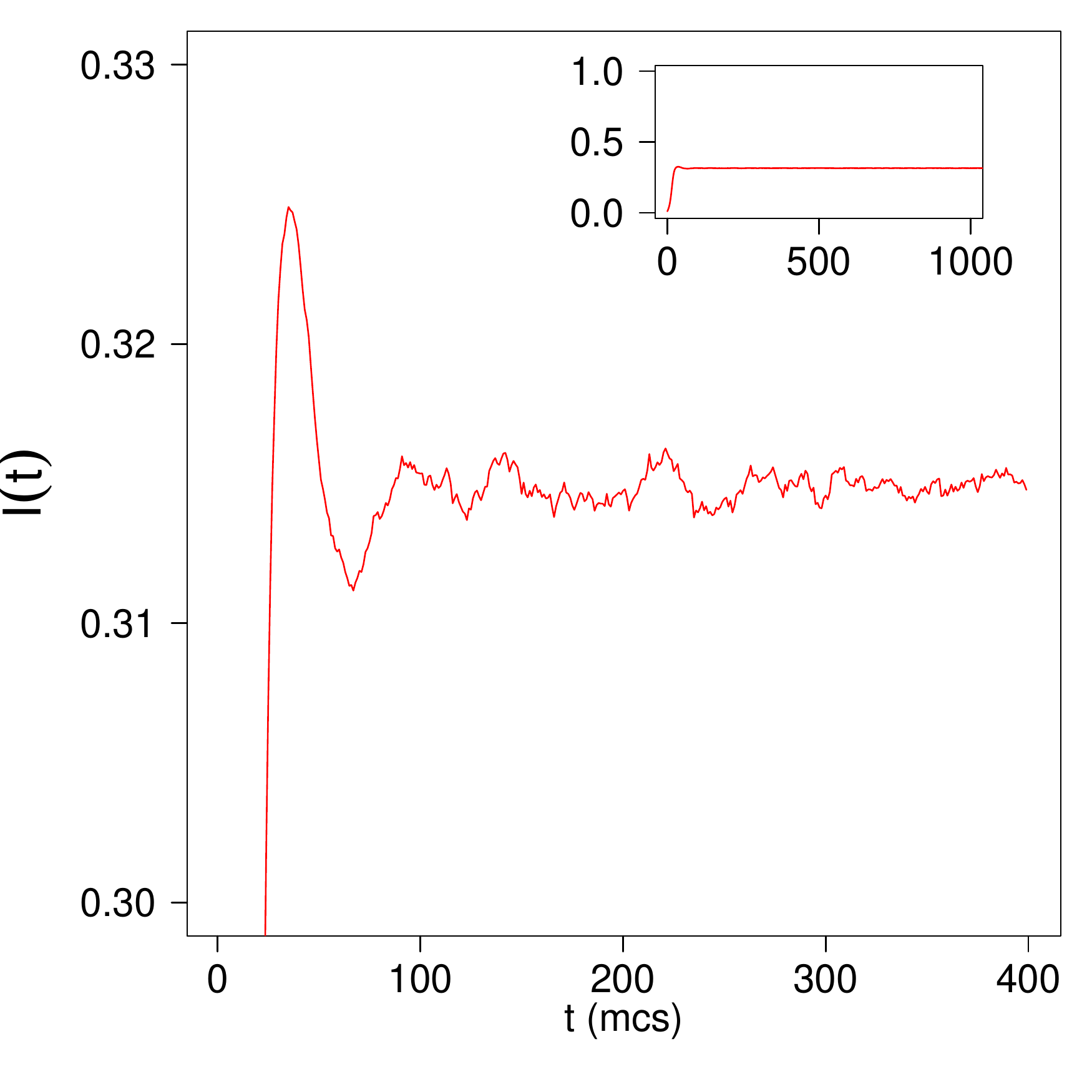}
            \end{subfigure}
      \begin{subfigure}[t]{0.30\textwidth}
        \includegraphics[width=\textwidth]{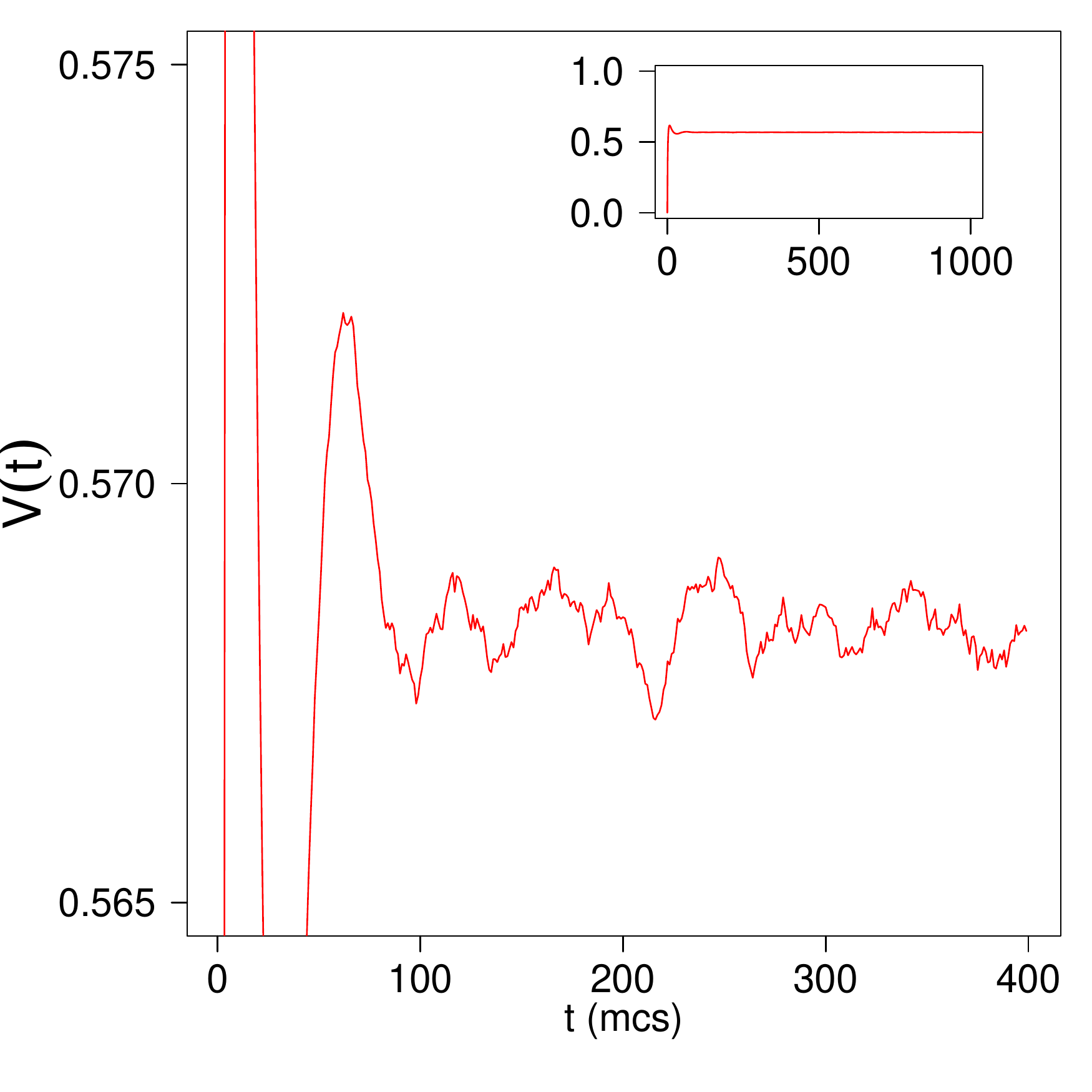}
            \end{subfigure}        
      \begin{subfigure}[t]{0.30\textwidth}
       \includegraphics[width=\textwidth]{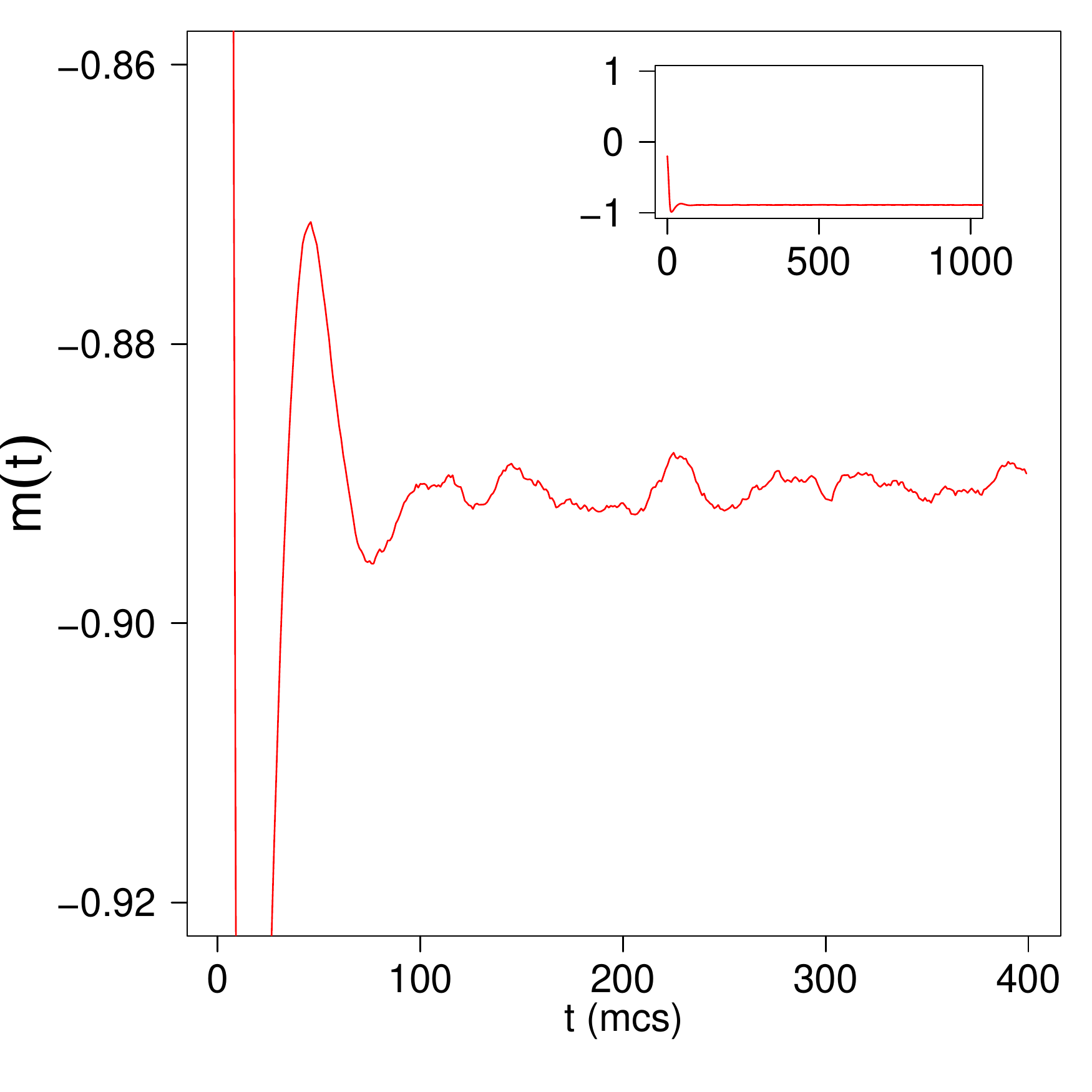}
            \end{subfigure}
    \end{center}

\caption{A closer look at the alternating temporal evolution of high and low values of $I(t)$, $V(t)$ and $m(t)$ for  $D=0.3$, $\lambda=0.9$,  $w=0.9$, $\alpha=0.1$,  $\phi=0.01$ and $N=10^{4}$. Here each subplot comes from Monte Carlo simulations taking the average over $100$ samples.}  
\label{fig:time-series-4}
\end{figure*}

In Fig. \ref{fig:time-series-4} we can see permanent alternations between high and low levels of vaccine coverage. These alternations arise from the competition between the social and epidemic pressures. This extends previous results for discrete opinion coupling that display oscillations such as  \cite{Voinson_Billiard_Alvergne2015} to continuous opinions. But, the cyclic dynamics presented here does not require a coexistence between positive and negative opinions as was the case in \cite{Voinson_Billiard_Alvergne2015}. 



\begin{figure*}[!ht]
\centering

\includegraphics[width=0.49\textwidth]{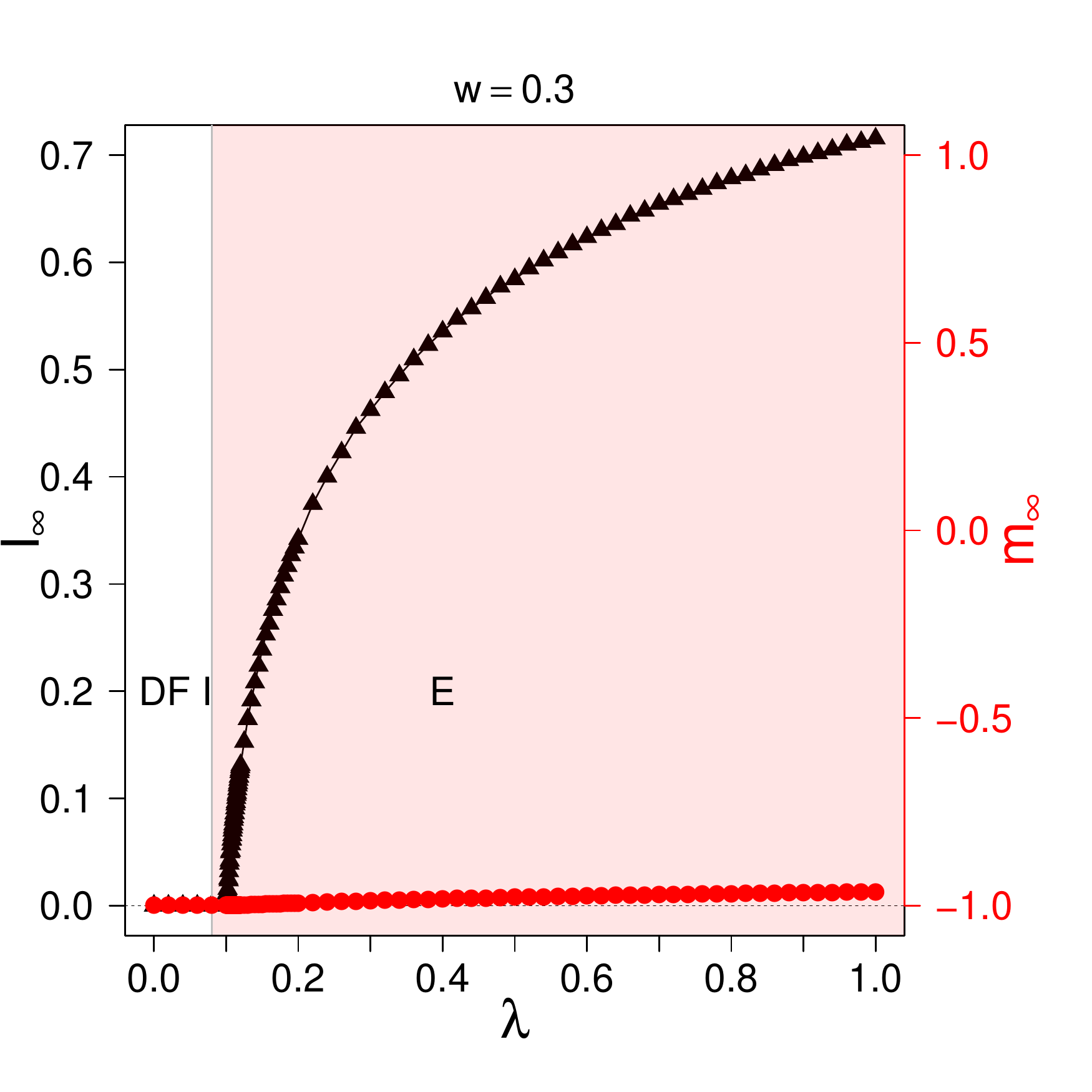}%
\includegraphics[width=0.49\textwidth]{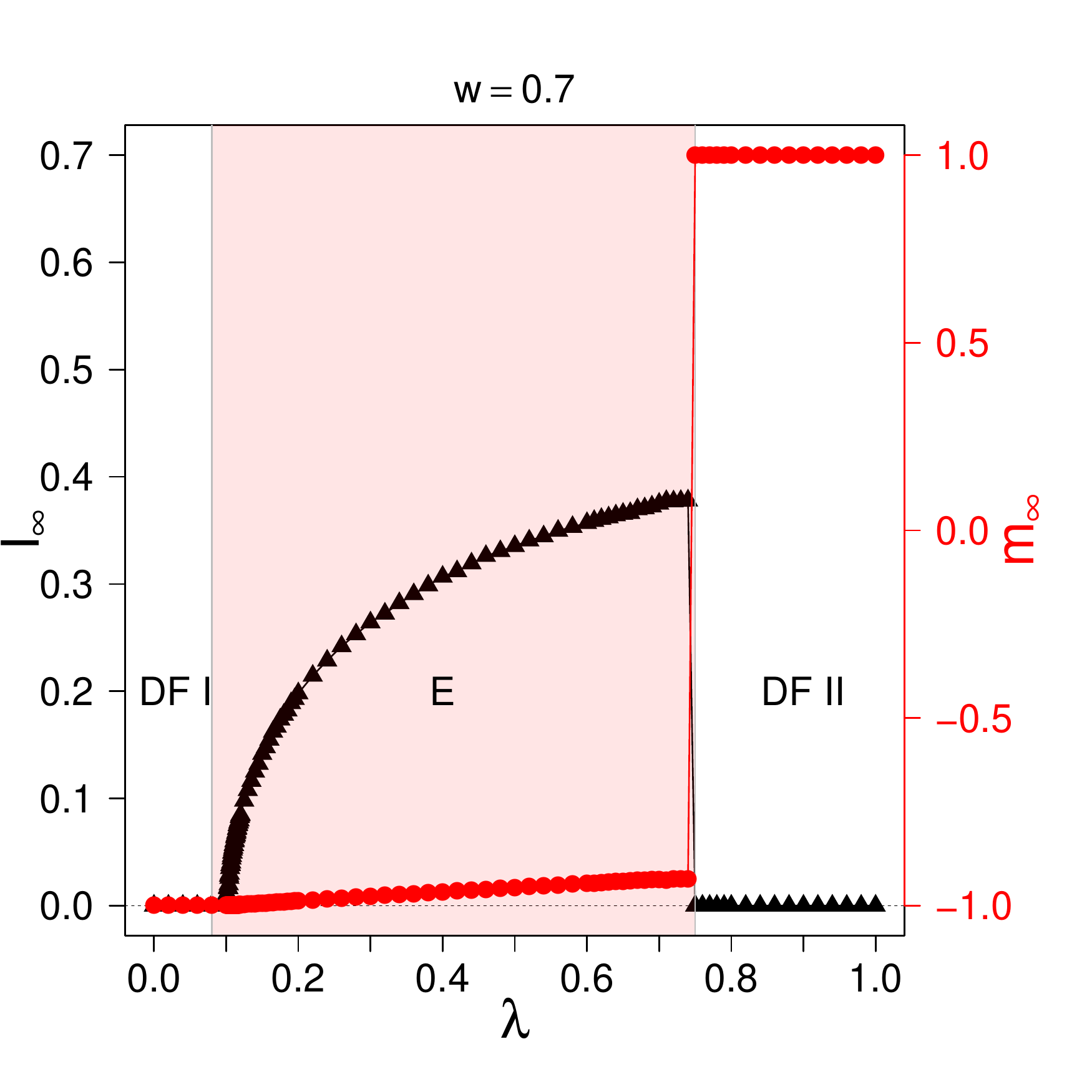}

\caption{Stationary density of Infected agents $I_{\infty}$ averaged only over surviving runs (left y-axis) and mean opinion $m_{\infty} =\sum_{i=1}^{N}o_i/N$  (right y-axis) as a function of $\lambda$ for $w=0.3$ (left) and $w=0.7$ (right). Parameters used are $D=0.1$, $\phi=0.01$, $\alpha=0.1$ and $N=10^{4}$. Data are averaged over $100$ independent simulations. Acronyms: \textbf{DF=Disease-Free, E=Endemic.} }
\label{fig:inf-vs-lambda-fullg1}
\end{figure*}

\begin{figure}[!ht]
\centering
            \includegraphics[width=0.49\textwidth]{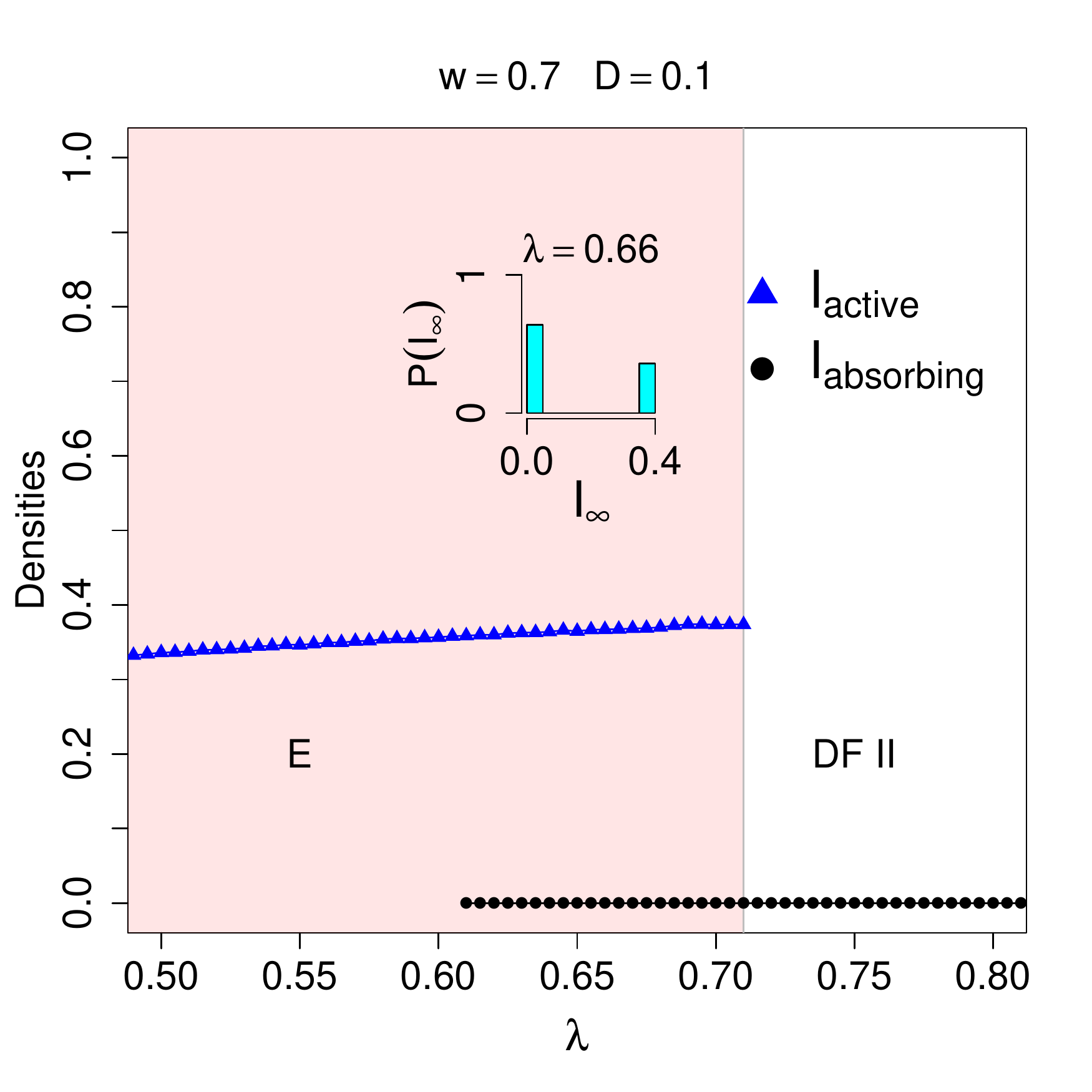}
            \includegraphics[width=0.49\textwidth]{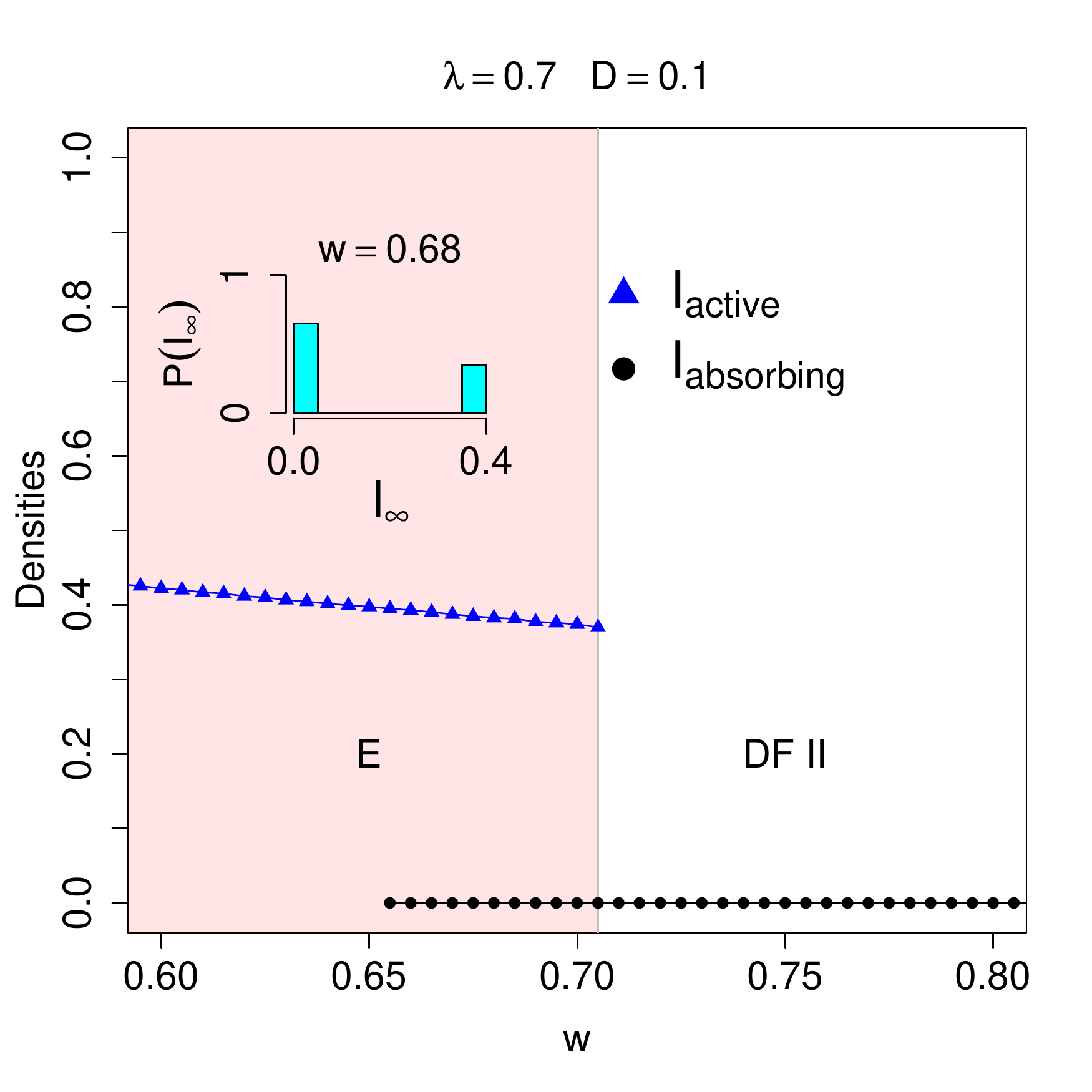}
        
            \includegraphics[width=0.49\textwidth]{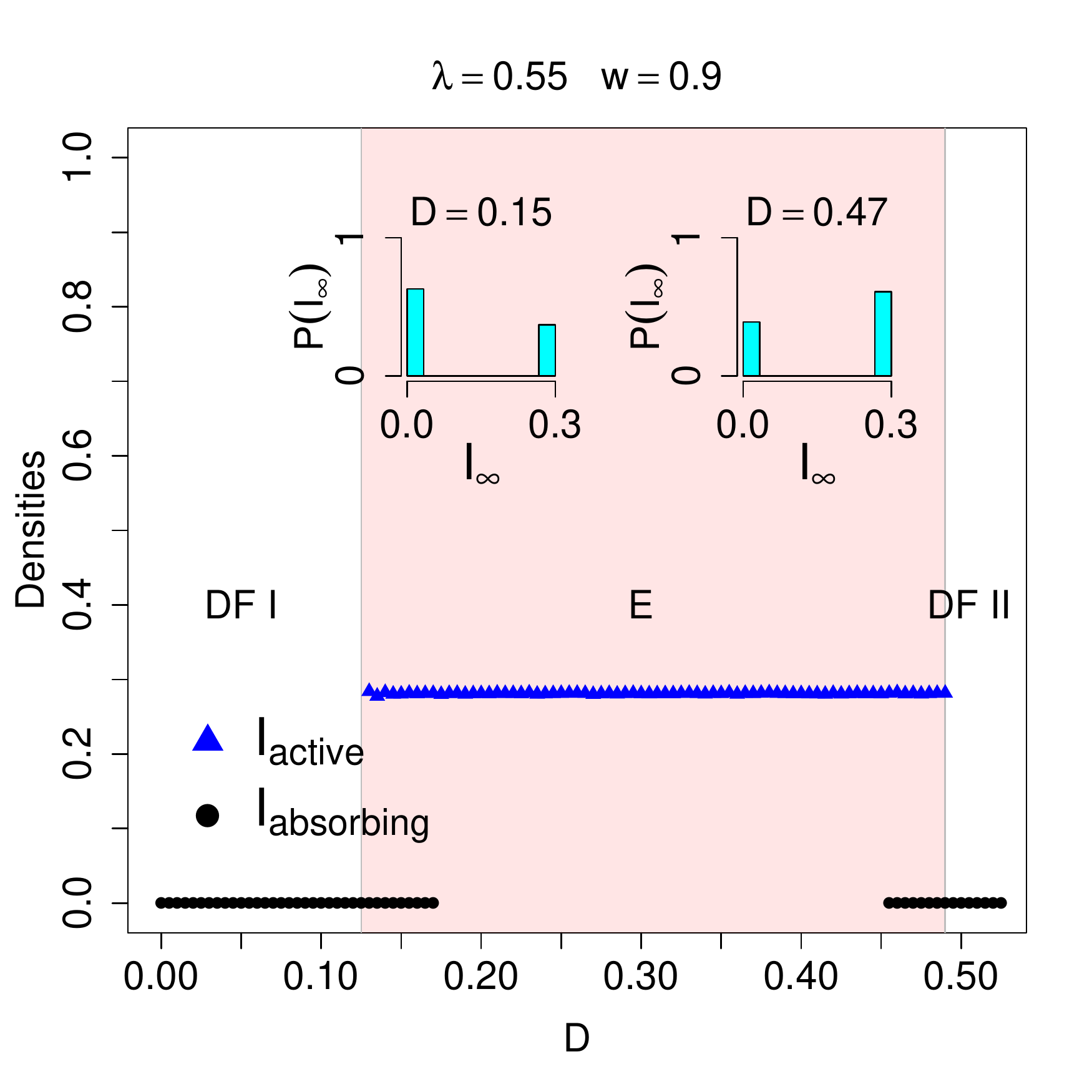}
            
    \caption{Stationary density of Infected agents $I_{active}$ averaged only over surviving   runs. 
    We use  $I_{absorbing}$ to indicate that there is at least one sample that has reached the absorbing 
    state (disease-free). Phase transition  for $\lambda$ (a) $w$ (b) and $D$ (c). We can observe that $I_{active}$ reveals four discontinuous phase transitions in our 
    coupled model. Parameters used here: $\phi=0.01$, $\alpha=0.1$, $N=10^{4}$. Data are averaged over $100$ independent simulations. Acronyms: \textbf{DF=Disease-Free, E=Endemic.}  }
\label{Fig:sudden-transtions}
\end{figure}

Fig. \ref{fig:inf-vs-lambda-fullg1} shows that the increment in the risk perception causes the appearance of a new first-order phase transition in $I_{\infty}$. For $w=0.3$ we can see that the usual continuous active-absorbing phase transition for $I_{\infty}$ is still present. This continuous phase transition arises purely due to epidemic reasons: imbalances between the disease spreading, immunization,  resusceptibility  and recovery. Very interestingly, for $w=0.7$ there is a second active-absorbing phase transition, but it is discontinuous. A more detailed analysis of this discontinuity, seen in Fig. \ref{Fig:sudden-transtions}, reveals the presence of the aforementioned bistability region. The insets show that the order-parameter probability distribution $P(I_{\infty})$ displays a  bimodal histogram. Such a coexistence between phases (active and absorbing in our case) is a signature of first-order phase transitions \cite{salinas,nrfim3d}. Curiously, tuning $w$ or $D$, as seen in Fig. \ref{Fig:sudden-transtions}, can also lead to sudden transitions.

The emergence of these new abrupt phase transitions is our main result. Abrupt phase transitions are not an odd outcome for social and biological contagion models \cite{Liu_Hethcote_Levin1987,Janssen_Muller_Stenull2004,Dodds_Watts2005,Gross_Lima_Blasius2006,Bagnoli_Lio_Sguanci2007,Bizhani_Paczuski_Grassberger2012,GomezGardenes2015,Cai2015,Chae_Yook_Kim2015,GomezGardenes2016,Liu2017,VelasquezRojas_Vazquez2017,cui2017}, but our work introduces a new mechanism that can lead to a first-order phase transition. Notice that we are using a SISV model that exhibits only a continuous phase transition. So, the discontinuity in $I_{\infty}$ in our coupled model is a consequence of the conflict between the social and epidemic pressures. Even though we could have used a vaccination dynamics that already undergoes a sudden transition \cite{KribsZaleta_VelascoHernandez2000}, this could hide the role played by the competition between peer pressure and risk perception.

In order to understand the underlying cause of the discontinuity in $I_{\infty}$ let us move back to equation $o_i(t)=o_i(t-1)+\epsilon o_j(t-1)+wI(t-1)$. In the absence of the external field  ($wI$) this equation always takes the opinions to one of the allowed extremes $o_i = \pm 1 $  in the stationary state, because of the bound imposed in the opinions. This bound to the opinions gives raise to somewhat weighted interactions. As an example consider the case in which $o_i = -0.9 \,\forall i$ and $wI = 0.3$. The opinion of any agent in the subsequent time step is given by $ o_i ' = -0.9 - \epsilon 0.9 + 0.3 $. As $\epsilon \in [0,1]$ the maximum possible decrement in the i-th opinion is of 0.1 for $\epsilon \geq 4/9$, and the maximum increment is of 0.3 when $\epsilon = 0$. This shows that the maximum increment is bigger than the maximum decrement. Therefore, even though the increment is less probable than the decrement it has a noticeable contribution. If the external field $wI$ is strong enough, as in Fig. \ref{fig:inf-vs-lambda-fullg1} for $w=0.7$, then the contribution of such increments lead the  agents' opinions to values slightly higher than $0$. This increment is already enough to start a shift in the overall opinion to $o_i = 1 \ \forall i $. This in turn leads to $\gamma_i=1 \ \forall i$, and that instantly  provokes the cessation of epidemic spreading since the effective
transmissibility is $(1-\gamma_i)\lambda I$.

\begin{figure*}[!ht]
\centering

    \begin{center}
        \begin{subfigure}[t]{0.49\textwidth}
            \includegraphics[width=\textwidth]{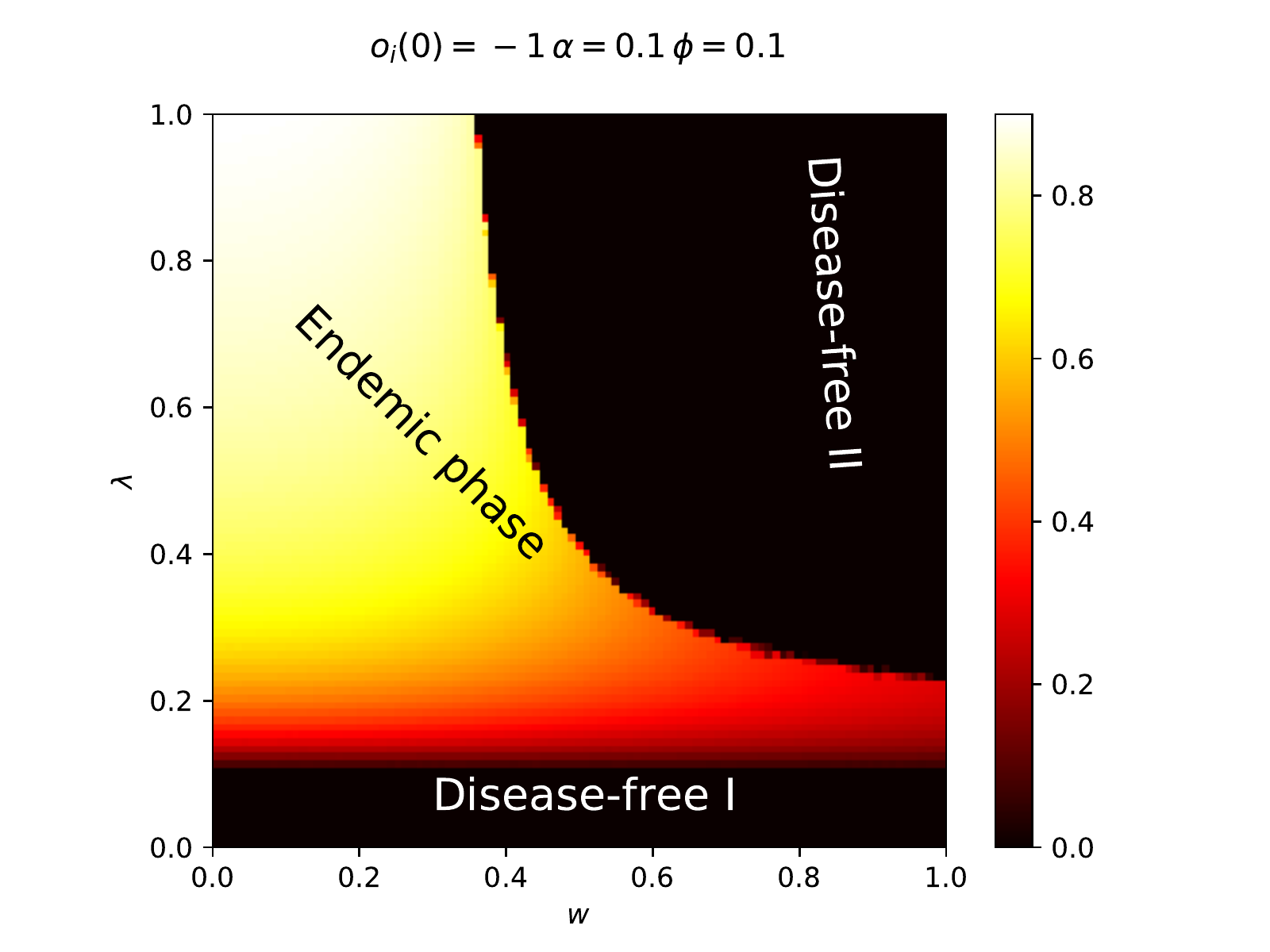}
            \caption{Initial condition: $o_i(t=0)=-1 \ \forall i$}
        \end{subfigure}
        \begin{subfigure}[t]{0.49\textwidth}
            \includegraphics[width=\textwidth]{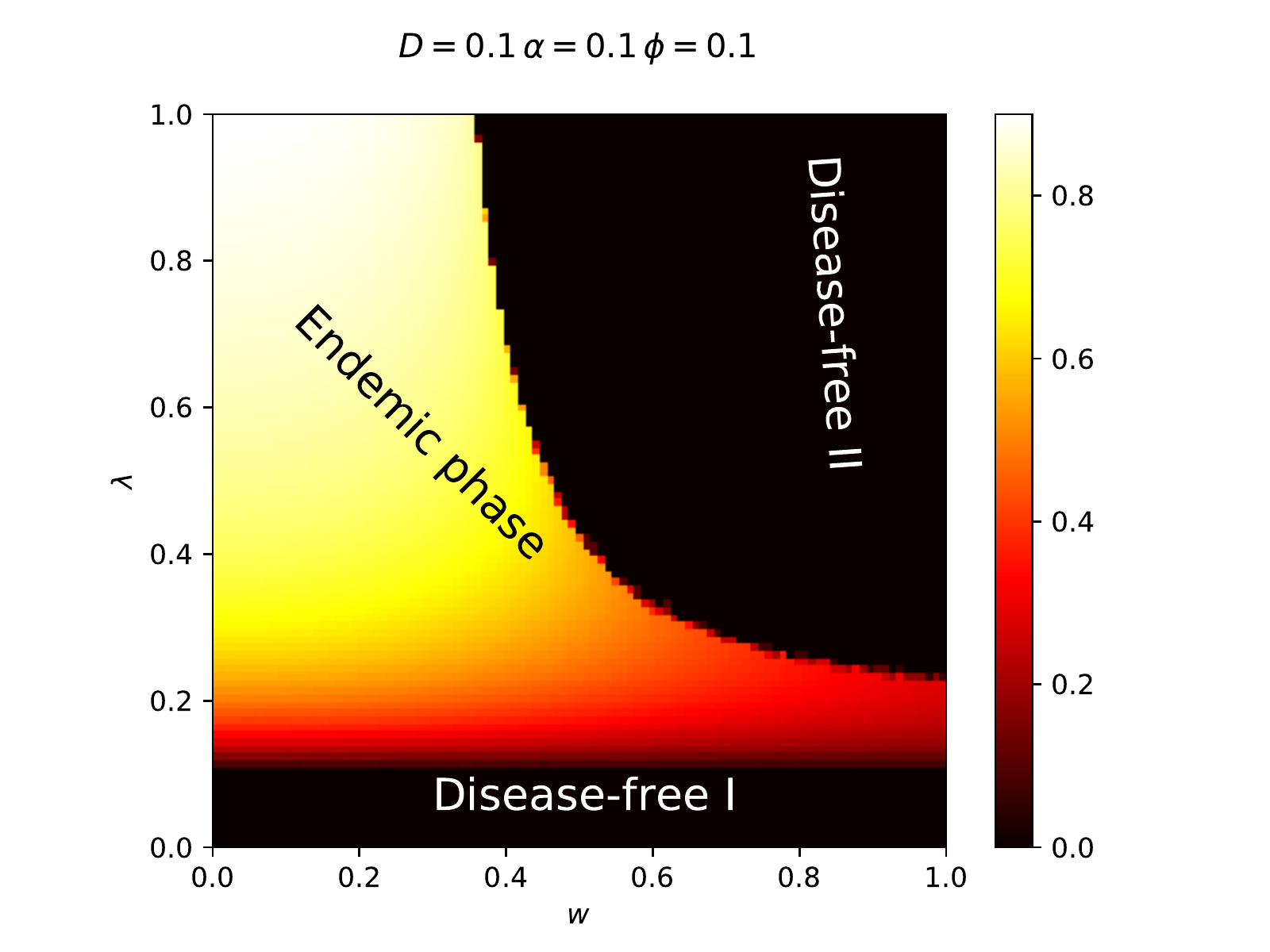}
            \caption{Initial condition: $D=10\%$}
        \end{subfigure} 
        \begin{subfigure}[t]{0.49\textwidth}
            \includegraphics[width=\textwidth]{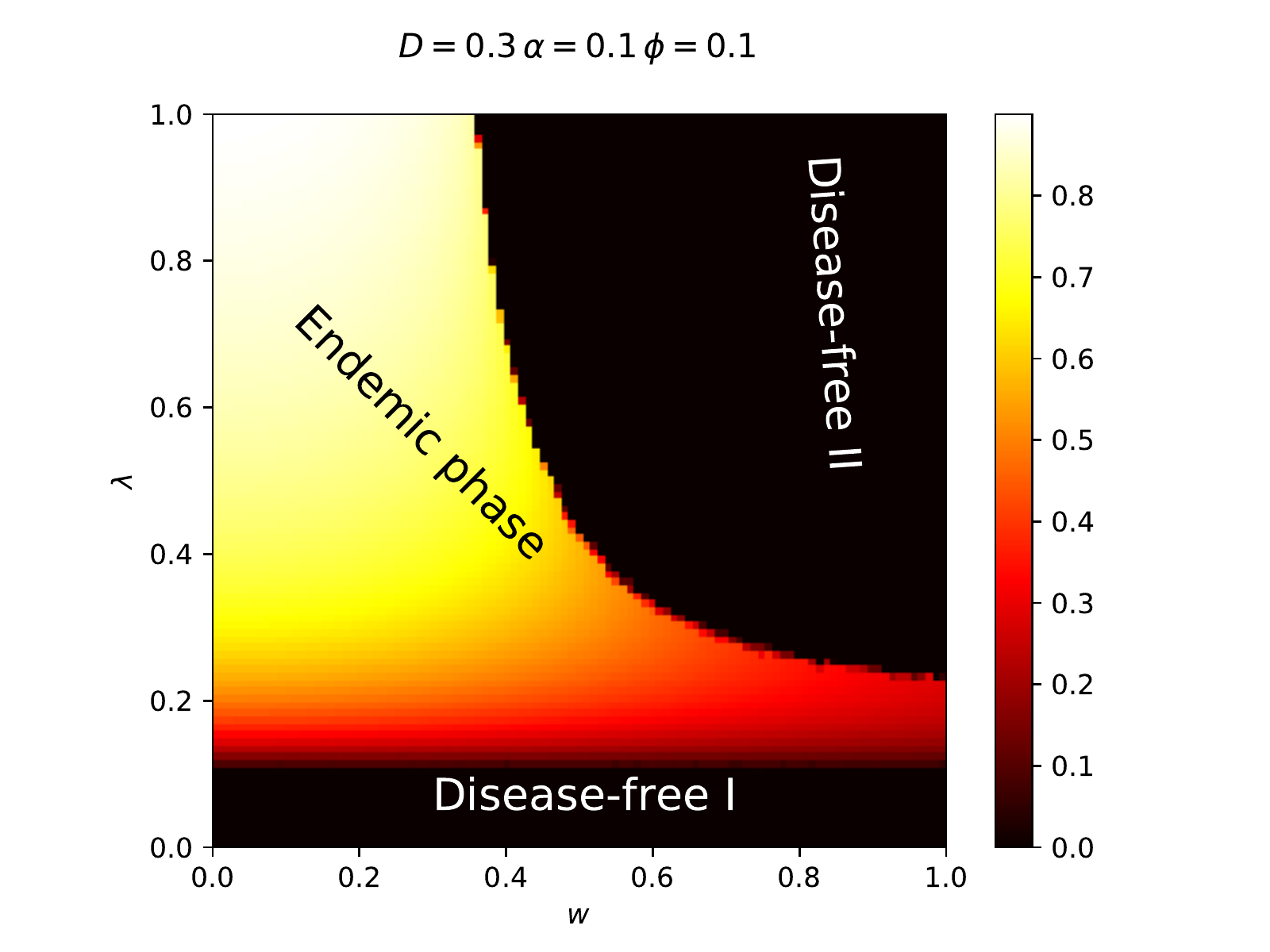}
            \caption{Initial condition: $D=30\%$}
        \end{subfigure}
        \begin{subfigure}[t]{0.49\textwidth}
            \includegraphics[width=\textwidth]{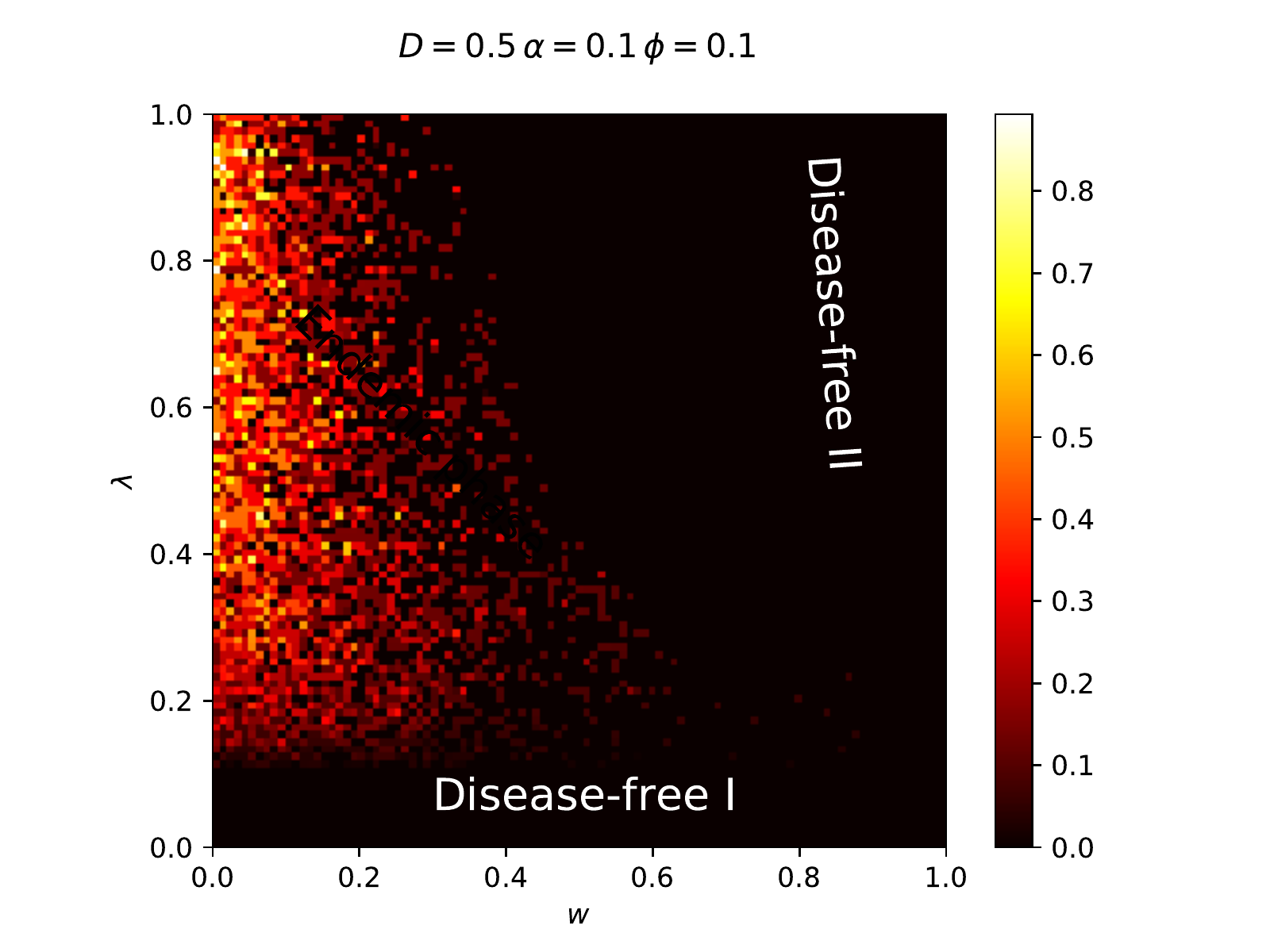}
            \caption{Initial condition: $D=50\%$}
        \end{subfigure}
    \end{center}

\caption{
Phase diagram of the stationary density of Infected $I_{\infty}$ agents as a function of $\lambda$ and $w$ for $\phi=0.10$. The population size is $N = 10^4$. For each parameter configuration (data point) $20$ samples were simulated, thus making a total of $200000$ independent simulations for each value of $D$. For panels (b-c-d) the initial configuration of the population is implemented as follows: $o_{i}(t=0) \sim U(0,1)$ with probability $D$ or $o_i(t=0) \sim  U(-1,0)$ with probability $1-D$, where $U(x_1,x_2)$ represents a random number uniformly distributed between $x_1$ and $x_2$.}  
\label{fig:PhaseDiagram-lam-x-w}
\end{figure*}

\begin{figure*}[!ht]
\centering

    \begin{center}
        \begin{subfigure}[t]{0.49\textwidth}
            \includegraphics[width=\textwidth]{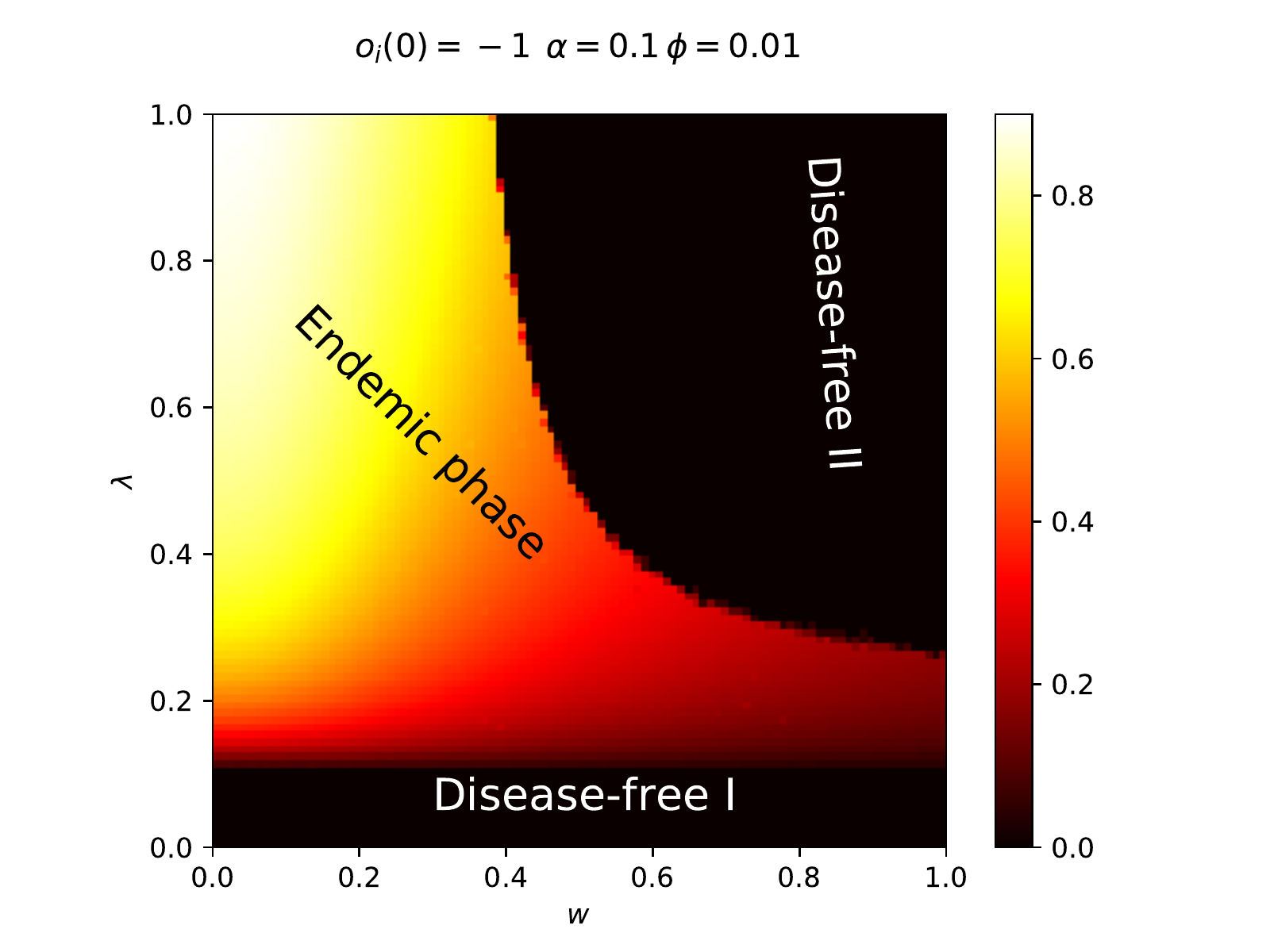}
            \caption{Initial condition: $o_i(t=0)=-1 \ \forall i$}
        \end{subfigure}
        \begin{subfigure}[t]{0.49\textwidth}
            \includegraphics[width=\textwidth]{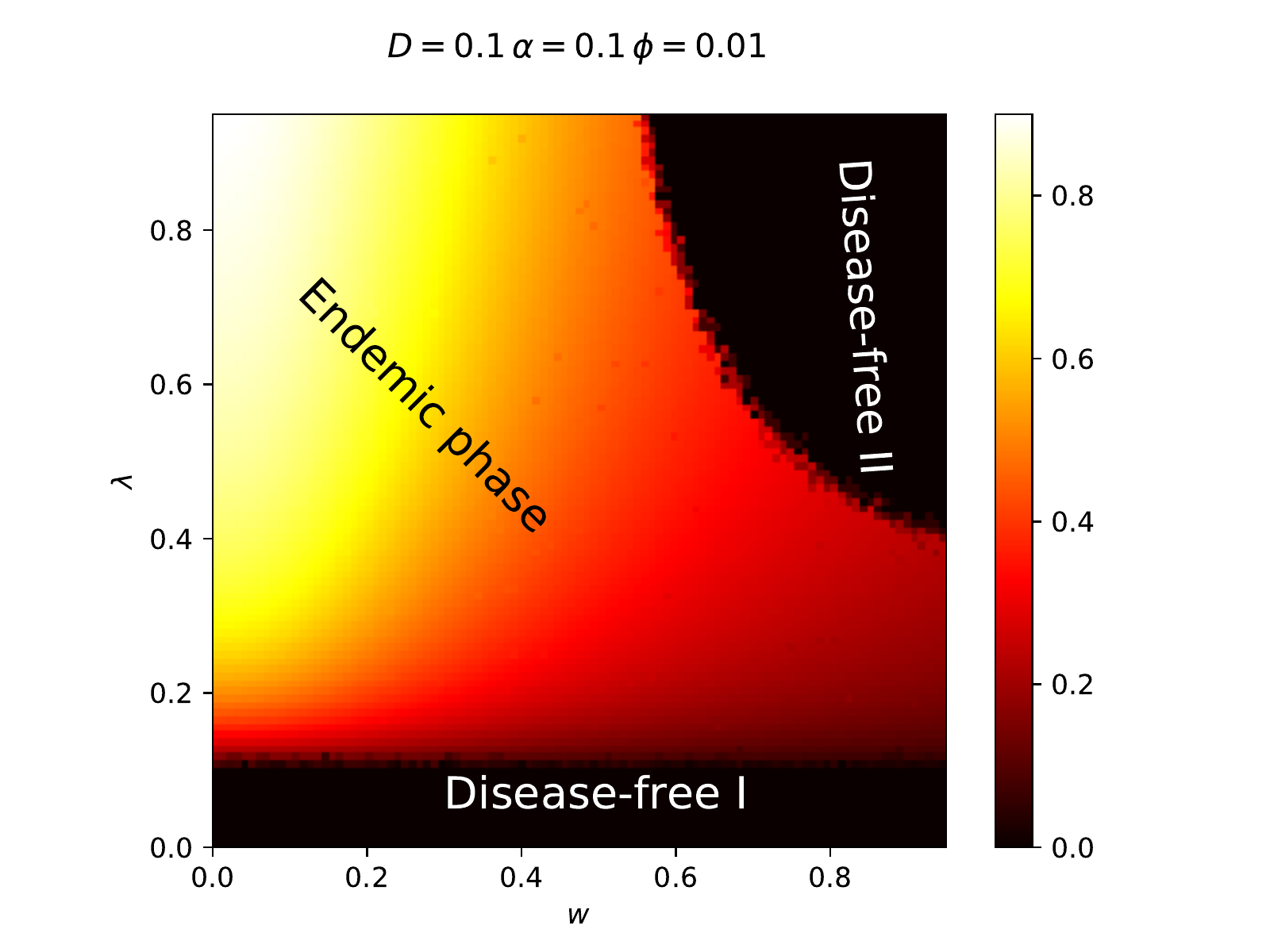}
            \caption{Initial condition: $D=10\%$}
        \end{subfigure}
        \begin{subfigure}[t]{0.49\textwidth}
            \includegraphics[width=\textwidth]{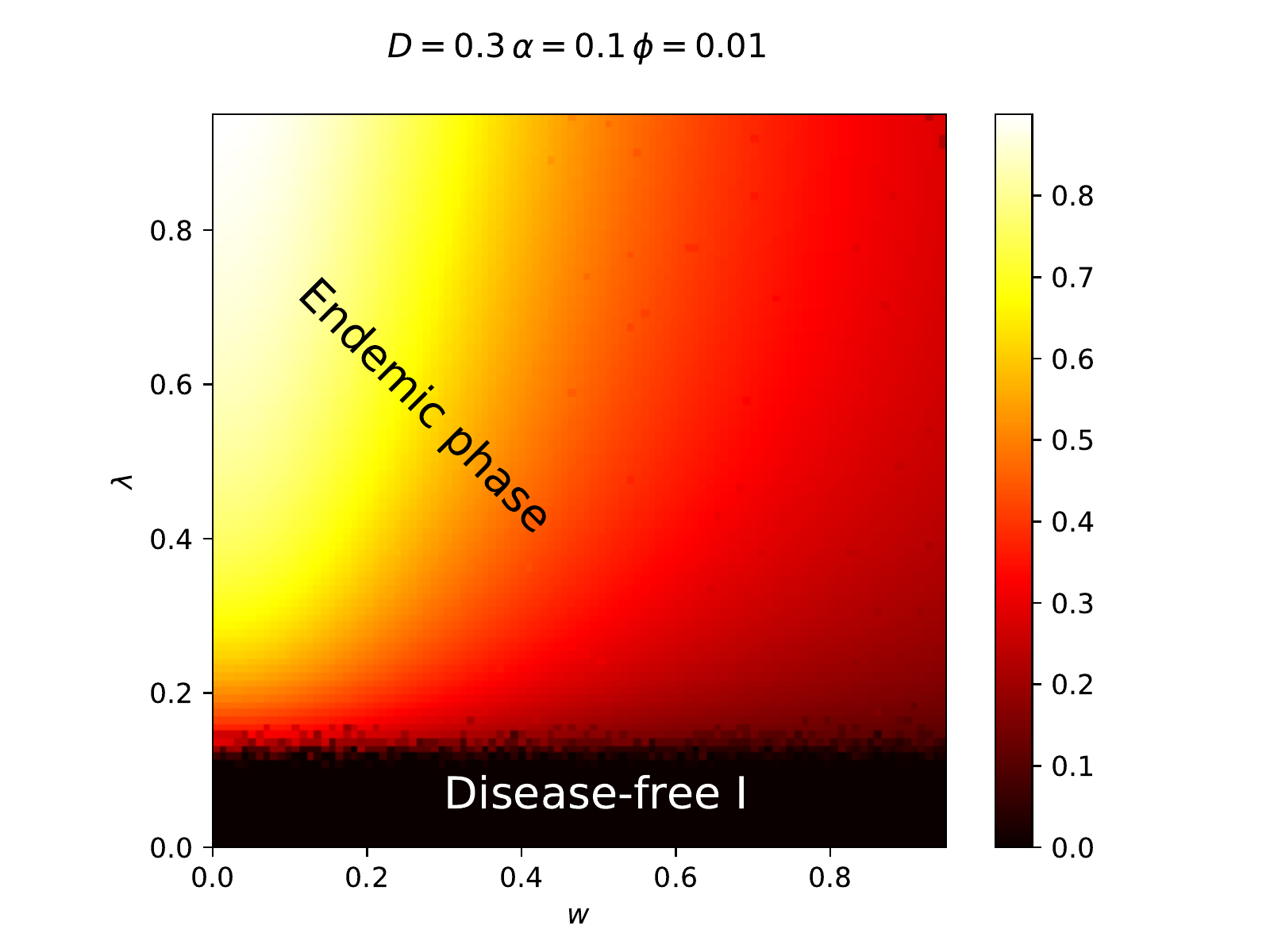}
            \caption{Initial condition: $D=30\%$}
        \end{subfigure}
        \begin{subfigure}[t]{0.49\textwidth}
            \includegraphics[width=\textwidth]{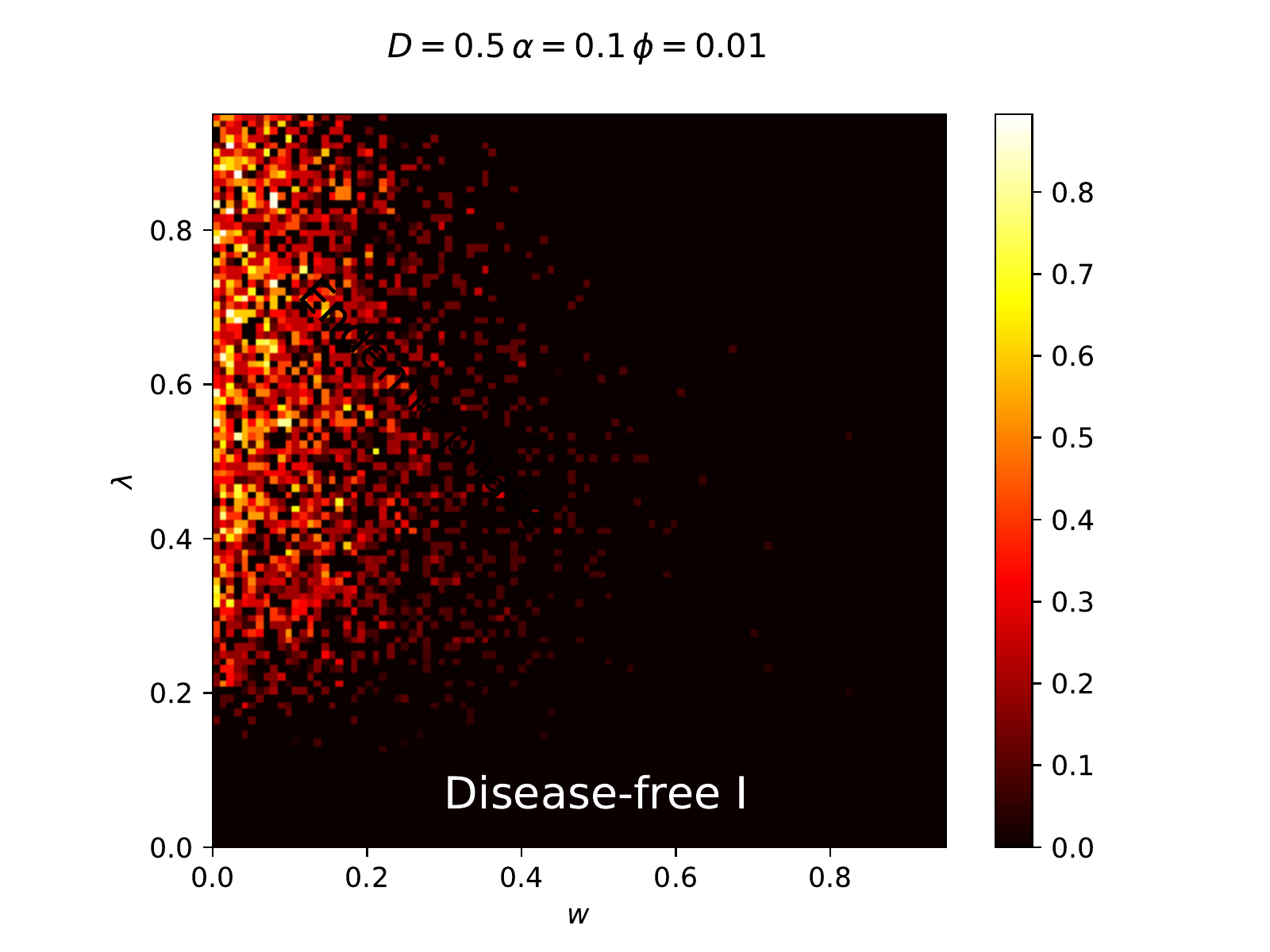}
            \caption{Initial condition: $D=50\%$}
        \end{subfigure}
    \end{center}

\caption{Same as Fig. \ref{fig:PhaseDiagram-lam-x-w}, except that now we have $\phi=0.01$. The effects of $D$ are more pronounced for $\phi=0.01$ than for $\phi=0.1$.}  
\label{fig:PhaseDiagram-lam-x-w-2}
\end{figure*}

\begin{figure*}[!ht]
\centering

    \begin{center}
        \begin{subfigure}[t]{0.49\textwidth}
            \includegraphics[width=\textwidth]{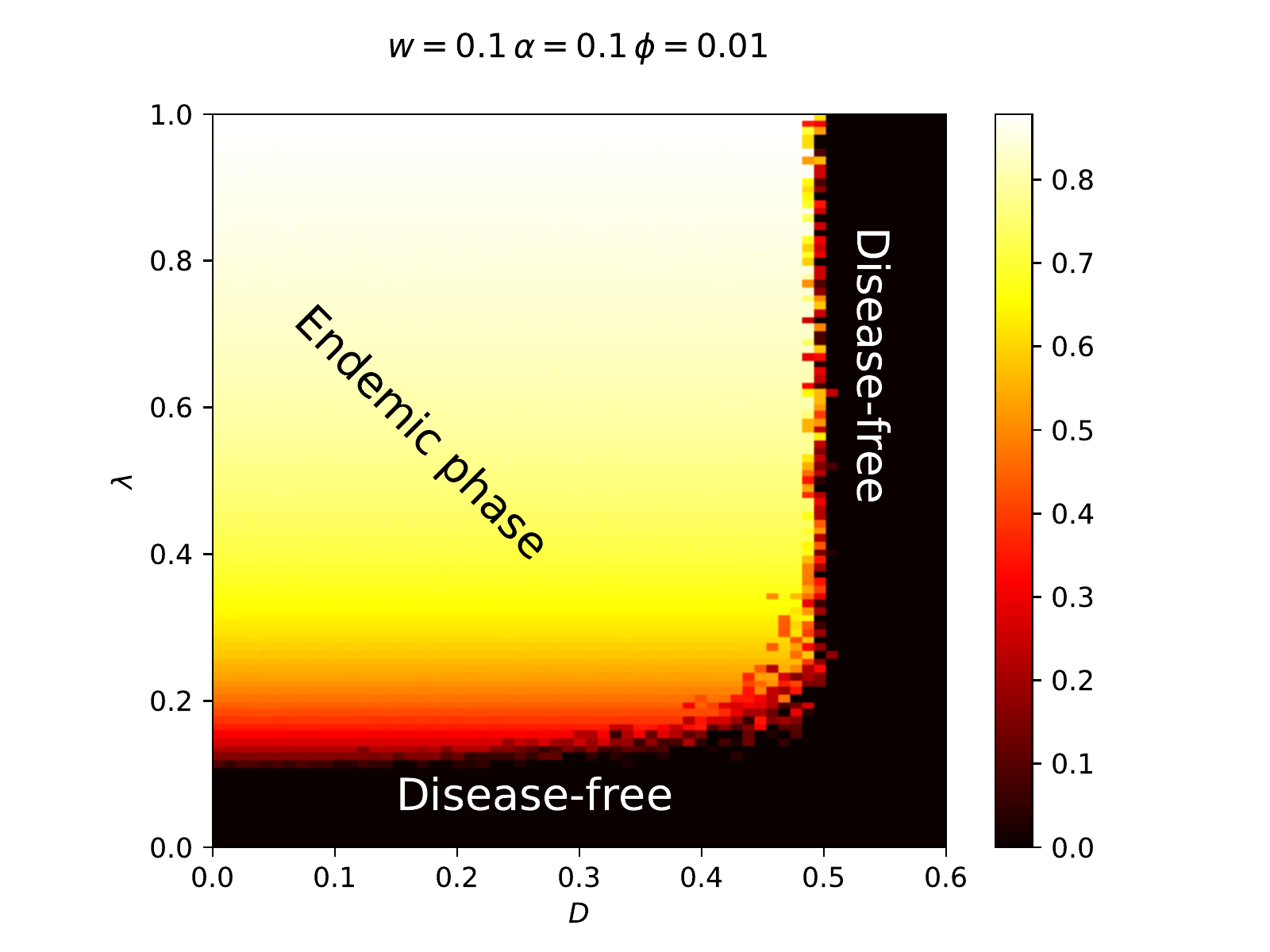}
            \caption{$w=0.1$}
        \end{subfigure}
        \begin{subfigure}[t]{0.49\textwidth}
            \includegraphics[width=\textwidth]{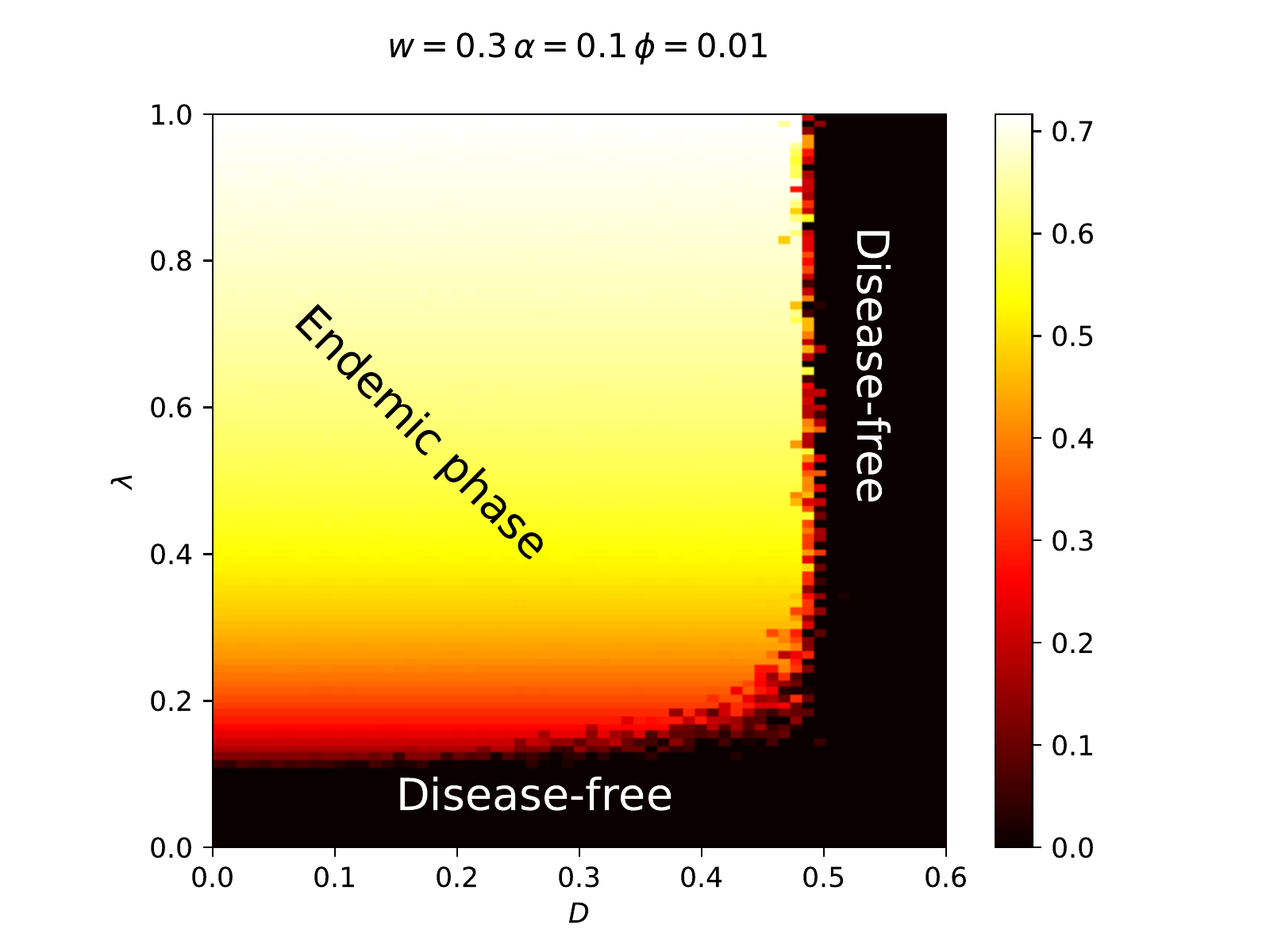}
            \caption{$w=0.3$}
        \end{subfigure}

        \begin{subfigure}[t]{0.49\textwidth}
            \includegraphics[width=\textwidth]{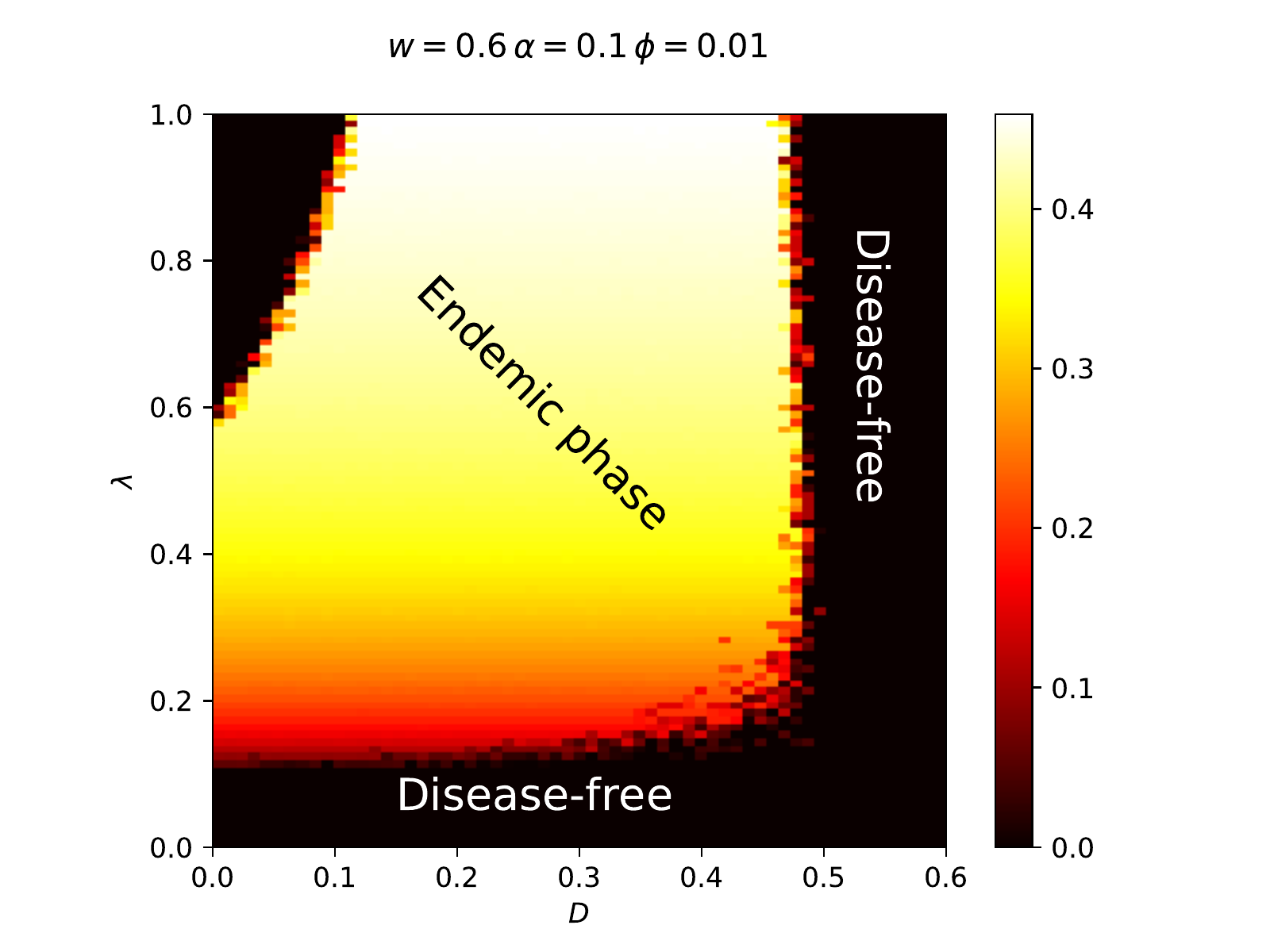}
            \caption{$w=0.6$}
        \end{subfigure}
        \begin{subfigure}[t]{0.49\textwidth}
            \includegraphics[width=\textwidth]{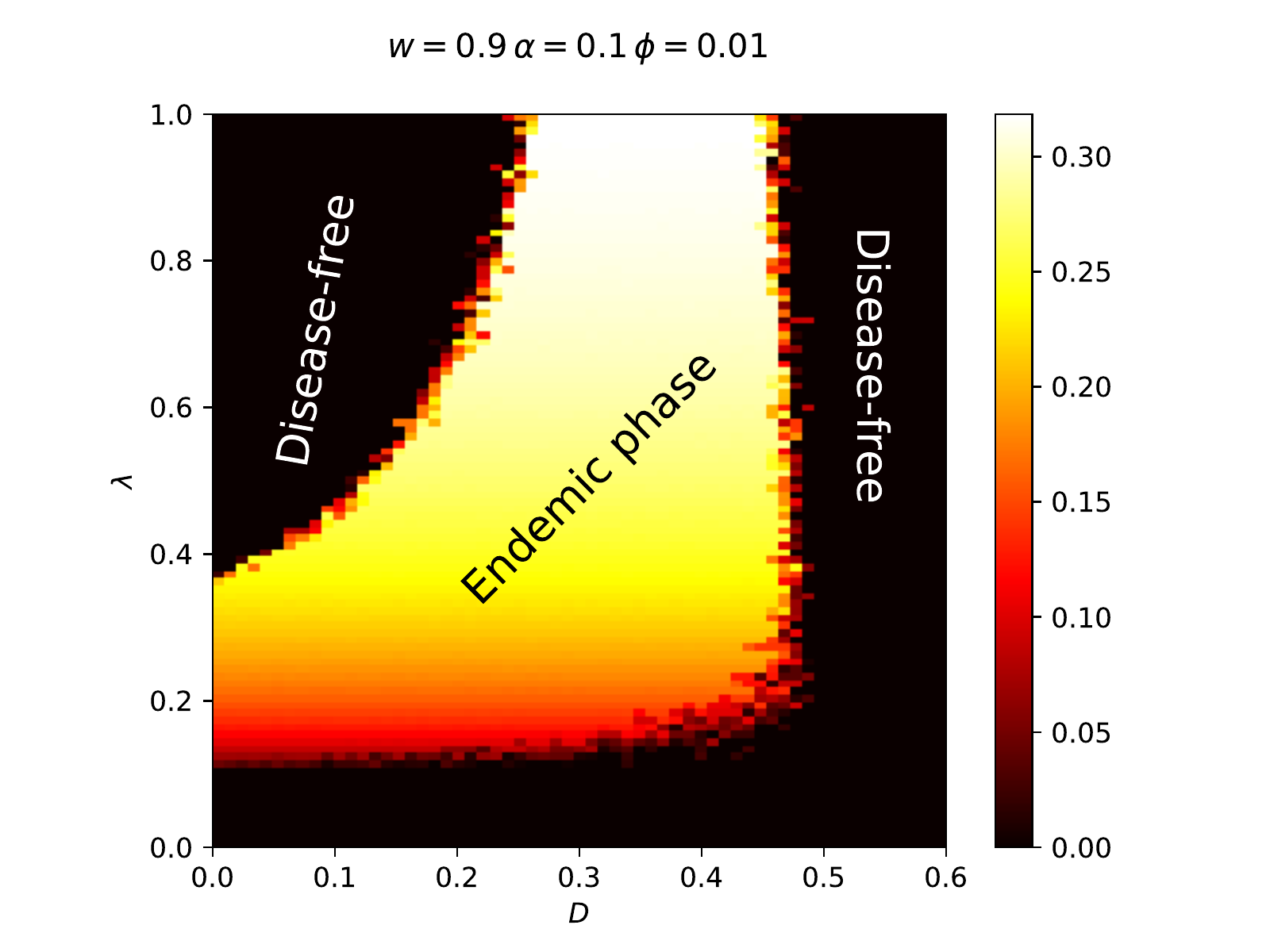}
            \caption{$w=0.9$}
        \end{subfigure}
    \end{center}

\caption{Phase diagram of the density of Infected agents $I_\infty$ versus $\lambda$ and $D$ for several values of $w$. Here we can see that $D > 0.5$ destroys the endemic state. The parameters were in $\lambda \in [0,1]$ and $D \in [0,0.6]$ both with a step of $0.01$. For each parameter configuration (data point) $20$ samples were simulated, thus making a total of $120000$ simulations for each value of $w$. }  
\label{fig:PhaseDiagram-lam-x-D-1}
\end{figure*}

A question immediately arises: how robust are these results? To address this we have performed extensive simulations of the model, as shown in Fig. \ref{fig:PhaseDiagram-lam-x-w}, \ref{fig:PhaseDiagram-lam-x-w-2} and \ref{fig:PhaseDiagram-lam-x-D-1}. These graphs show that our results hold for a large set of parameter values. Notice that the onset of the endemic phase does not depend on neither $w$ or $D$. This was expected, since before the onset there is no stationary disease transmission that could spread fear in the population. This leads us to conclude that the coupling between social and epidemic processes does not intervene in the transition $DF I \rightarrow E$. Observe also that there are combinations of parameter values that suppress the transition $E \rightarrow DF II$.

\clearpage

\clearpage

\clearpage

\clearpage

\section{Final Remarks}

We have studied a dynamical model considering a coupling of an epidemic spreading and an opinion dynamics. The epidemic dynamics is based on the SIS model with an extra Vaccinated compartment, whereas the opinions are ruled by a kinetic continous opinion model. In this case, we consider a competition among individuals with positive and negative opinions about vaccination.

\vspace{-0.7cm}

We verified through extensive numerical simulations that, in addition to the usual continuous phase transition of SIS-like models, our system can undergo discontinuous transitions among active (endemic) and absorbing (disease free) steady states. The emergence of these new abrupt phase transitions is our main result. From a theoretical viewpoint our contribution is in the introduction of a new mechanism that generates a first-order phase transition in nonequilibrium systems, namely a competition between social pressure regarding vaccination and epidemic evolution. From the practical point of view, our results suggest that an increment in the initial fraction of pro-vaccine individuals has a twofold effect: it can lead to smaller epidemic outbreaks in the short term, but it also contributes to the survival of the chain of infections in the long term. This is a counterintuitive outcome, but it is in line with empirical observations that vaccines can become a victim of their own success.

\vspace{-0.18cm}

There are some open questions that we intend to address in the near future. First, it would be  interesting to extend the equation $o_i(t+1)=o_i(t)+\epsilon o_j(t)+wI(t)$ to a networked population in order to study  how the abrupt phase transitions observed here change if the perception of the risk of being infected depends on the fraction of infected neighbors (local information) \cite{Bagnoli_Lio_Sguanci2007}. Second, in this work all agents have the same degree of conviction, then another highly promising avenue of research is to consider a heterogeneous power of conviction for the agents. Indeed, as discussed in \cite{Galam2010}, agents on the refusal side (anti-vaccine) are more convicted of their opinions than pro-vaccine agents. This line of research will allow us to evaluate the impact of recommendations for vaccination from  agents with naturally high social status such as health care providers which is an important factor for vaccination adherence \cite{Dorell2013}. This is an issue that  has not yet been addressed in the literature of coupled vaccination-opinion dynamics  \cite{Verelst_Willem_Beutels2016}.


\section*{Acknowledgments}

The authors acknowledge financial support from the Brazilian funding agencies Conselho Nacional de Desenvolvimento Cient\'ifico e Tecnol\'ogico (CNPq), Coordena\c{c}\~ao de Aperfei\c{c}oamento de Pessoal de N\'ivel Superior (CAPES) and Funda\c{c}\~ao Carlos Chagas Filho de Amparo \`a Pesquisa do Estado do Rio de Janeiro (FAPERJ). The authors also thank Silvio M. D. Queir\'os for a critical reading of the manuscript.

\appendix





\clearpage

\section{Only opinion model with external field}

In order to understand the mechanisms of the model it is useful to analyse its individual parts. In this case, we consider in this appendix the opinion model separated from the epidemic model. Thus, we get a simple opinion model with an external field. Albeit, this simple model does not account for the whole results it sheds some light in the mechanism behind the unusual discontinuous phase transition.

\begin{figure*}[!ht]
\centering
\includegraphics[width=0.49\textwidth]{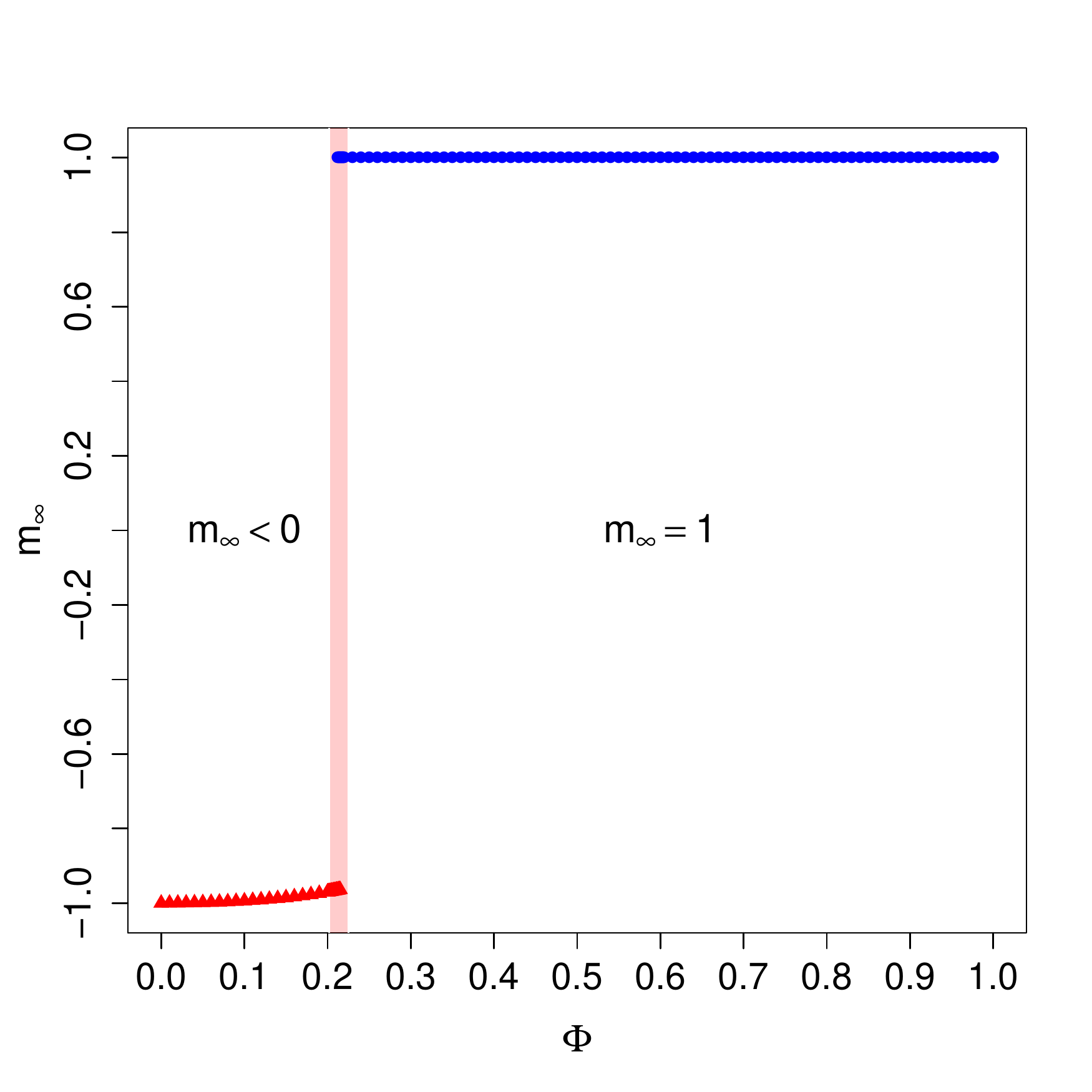}%

    \caption{The order parameter $m_{\infty} $ versus the external field $\Phi$.  Simulations were performed for a population size $N=10^4$ agents over $100$ samples and the starting configurations consider $D=0$. Here we have a bistability region (shaded) for $\Phi \approx [0.212,0.215]$, where an abrupt transition can happen.}

\label{Fig:OxPhi}    
\end{figure*}

With this in mind the simple opinion model is governed by the following equation

\begin{equation}
    o_i (t+1) = o_i (t) + \epsilon o_j (t) + \Phi ~,
\end{equation}

\noindent
where $\Phi$ is the external field. For the purpose of our model the external field belongs to the interval $\Phi \in [0 ,1]$. From Fig. 
\ref{Fig:OxPhi} 
 we can see that the mean opinion $m_{\infty} =\sum_{i=1}^{N}o_i/N$ suffers an abrupt change for external fields $\Phi \approx [0.212,0.215]$. This abrupt change in the opinions is the underlying cause for the abrupt phase transitions that appeared in the coupled model presented in this paper.

\clearpage

\bibliographystyle{plain}

\begin{thebibliography}{99}


\bibitem{Wang2016}
Z. Wang \textit{et. al.}, ``Statistical physics of vaccination'', Physics Reports 664 (2016): 1-113.


\bibitem{Satorras2015}
R. Pastor-Satorras \textit{et. al.}, ``Epidemic processes in complex networks'', Reviews of modern physics 87.3 (2015): 925.


\bibitem{Castellano2009}
C. Castellano, S. Fortunato, V. Loreto, ``Statistical physics of social dynamics'', Reviews of modern physics 81.2 (2009): 591.

\bibitem{Galam2012}
S. Galam, Sociophysics: A Physicist's Modeling of Psycho-Political Phenomena (Springer, Berlin, 2012).


\bibitem{Sen2013}
P. Sen, B. K. Chakrabarti, Sociophysics: An Introduction (Oxford University Press, Oxford, 2013).



\bibitem{Moreno2002}
Y. Moreno, R. Pastor-Satorras, A. Vespignani, ``Epidemic outbreaks in complex heterogeneous networks'', The European Physical Journal B - Condensed Matter and Complex Systems 26.4 (2002): 521-529.


\bibitem{Satorras2001}
R. Pastor-Satorras, A. Vespignani, ``Epidemic dynamics and endemic states in complex networks'', Physical Review E 63.6 (2001): 066117.



\bibitem{CrokidakisMenezes2012}
N. Crokidakis, M. A. de Menezes, ``Critical behavior of the SIS epidemic model with time-dependent infection rate'', Journal of Statistical Mechanics: Theory and Experiment 2012.05 (2012): P05012.


\bibitem{Antulov2015}
N. Antulov-Fantulin \textit{et. al.}, ``Identification of Patient Zero in Static and Temporal Networks: Robustness and Limitations'', Physical Review Letters 114.24 (2015): 248701.

\bibitem{Kitsak2010}
M. Kitsak \textit{et. al.}, ``Identification of influential spreaders in complex networks'', Nature physics 6.11 (2010): 888-893.

\bibitem{Xiong2017}
F. Xiong, Zhao-Yi Li, ``Effective Methods of Restraining Diffusion in Terms of Epidemic Dynamics'', Scientific Reports 7 (2013) 6013.

\bibitem{Galam2010}
S. Galam, ``Public debates driven by incomplete scientific data: The cases of evolution theory, global warming and H1N1 pandemic influenza'', Physica A: Statistical Mechanics and its Applications 389.17 (2010): 3619-3631.

\bibitem{Salathe_Bonhoeffer2008}
M. Salath\'{e}, S. Bonhoeffer, ``The effect of opinion clustering on disease outbreaks'', Journal of The Royal Society Interface 5.29 (2008): 1505-1508.

\bibitem{Wang2015}
Z. Wang \textit{et. al.}, ``Coupled disease-behavior dynamics on complex networks: A review'', Physics of life reviews 15 (2015): 1-29.

\bibitem{Verelst_Willem_Beutels2016}
F. Verelst, L. Willem, P. Beutels, ``Behavioural change models for infectious disease transmission: a systematic review (2010–2015)'', Journal of The Royal Society Interface 13.125 (2016): 20160820.

\bibitem{Funk2010}
S. Funk, M. Salath\'{e} and V.A.A. Jansen, ``Modelling the influence of human behaviour on the spread of infectious diseases: a review'', Journal of the Royal Society Interface 7.50 (2010): 1247-1256.

\bibitem{allan_celia_nuno}
A. R. Vieira, C. Anteneodo, N. Crokidakis, ``Consequences of nonconformist behaviors in a continuous opinion model'', Journal of Statistical Mechanics: Theory and Experiment 023204 (2016).

\bibitem{PiresCrokidakis2017}
M. A. Pires, N. Crokidakis, ``Dynamics of epidemic spreading with vaccination: Impact of social pressure and engagement'', Physica A: Statistical Mechanics and its Applications 467 (2017): 167-179.


\bibitem{Zhou_Yicang_Liu2003}
Y. Zhou, H. Liu, ``Stability of periodic solutions for an SIS model with pulse vaccination'', Mathematical and Computer Modelling 38.3 (2003): 299-308.

\bibitem{Shaw2010}
L. B. Shaw, I. B. Schwartz, ``Enhanced vaccine control of epidemics in adaptive networks'', Physical Review E 81.4 (2010): 046120.

\bibitem{CoelhoCodeco2009}
F. C. Coelho, C. T. Code\c{c}o, ``Dynamic modeling of vaccinating behavior as a function of individual beliefs'', PLoS Comput Biol 5.7 (2009): e1000425.

\bibitem{Eames2009}
K. T. D. Eames. ``Networks of influence and infection: parental choices and childhood disease'', Journal of The Royal Society Interface 6.38 (2009): 811-814.

\bibitem{Voinson_Billiard_Alvergne2015}
M. Voinson, S. Billiard, A. Alvergne, ``Beyond Rational Decision-Making: Modelling the Influence of Cognitive Biases on the Dynamics of Vaccination Coverage'', PloS one 10.11 (2015): e0142990.


\bibitem{Zuzek2017}
L. G. Alvarez-Zuzek \textit{et. al.}, ``Epidemic spreading in multiplex networks influenced by opinion exchanges on vaccination'', PLoS ONE 12(11):e0186492 (2017).


\bibitem{Xia2013}
S. Xia, J. Liu, ``A computational approach to characterizing the impact of social influence on individuals- vaccination decision making'', PLoS One 8.4 (2013): e60373.


\bibitem{Lau2010}
J. T. F. Lau \textit{et. al.}, ``Factors in association with acceptability of A/H1N1 vaccination during the influenza A/H1N1 pandemic phase in the Hong Kong general population'', Vaccine 28.29 (2010): 4632-4637.

\bibitem{Bish2011}
A. Bish \textit{et. al.}, ``Factors associated with uptake of vaccination against pandemic influenza: a systematic review'',\,\, Vaccine 29.38 (2011): 6472-6484.

\bibitem{LCCC2010}
M. Lallouache \textit{et. al.}, ``Opinion formation in kinetic exchange models: Spontaneous symmetry-breaking transition'', Physical Review E 82.5 (2010): 056112.

\bibitem{Cava2005}
M. A. Cava \textit{et. al.}, ``Risk perception and compliance with quarantine during the SARS outbreak'', Journal of Nursing Scholarship 37.4 (2005): 343-347.

\bibitem{Bottcher2015}
 L. Bottcher \textit{et. al.}, ``Disease-induced resource constraints can trigger explosive epidemics'', Scientific Reports 5 (2015) 16571.

\bibitem{Conte2012}
R. Conte \textit{et al.}, ``Manifesto of computational social science'', The European Physical Journal Special Topics 214.1 (2012): 325-346.


\bibitem{Wu2013}
Z. X. Wu, H. F. Zhang, ``Peer pressure is a double-edged sword in vaccination dynamics'', Europhysics Letters 104.1 (2013): 10002.

\bibitem{Maharaj_Kleczkowski2012}
S. Maharaj, A. Kleczkowski, ``Controlling epidemic spread by social distancing: Do it well or not at all'', BMC Public Health 12.1 (2012): 679.

\bibitem{CrokidakisQueiros2012}
N. Crokidakis, S. M. D. Queir\'os, ``Probing into the effectiveness of self-isolation policies in epidemic control'', Journal of Statistical Mechanics: Theory and Experiment 2012.06 (2012): P06003.

\bibitem{Zhang2013}
H. F. Zhang \textit{et. al.}, ``Braess's Paradox in Epidemic Game: Better Condition Results in Less Payoff'', Scientific Reports 3 (2013) 3292.
  

\bibitem{Zhang2017}
H. F. Zhang \textit{et. al.}, ``Preferential imitation can invalidate targeted subsidy policies on seasonal-influenza diseases'', Applied Mathematics and Computation 294 (2017): 332-342.
  
  
  
\bibitem{LeeMale2011}
M. S. W. Lee, M. Male, ``Against medical advice: the anti‐consumption of vaccines'', Journal of Consumer Marketing, Vol. 28 Issue: 7, pp.484-490 (2011).

\bibitem{Larson2011}
H. J. Larson \textit{et. al.}, ``Addressing the vaccine confidence gap'', The Lancet 378.9790 (2011): 526-535.


\bibitem{salinas}
S.-Ho Tsai, S. R. Salinas, ``Fourth-Order Cumulants to Characterize the Phase Transitions of a Spin-1 Ising Model'', Braz. J. Phys. 28 (1998) 587.


\bibitem{nrfim3d}
N. Crokidakis, ``Nonequilibrium phase transitions and tricriticality in a three-dimensional lattice system with random-field competing kinetics'', Physical Review E 81 (2010) 041138.






%

\bibitem{Liu_Hethcote_Levin1987}
W. Liu, H. W. Hethcote, S. A. Levin, ``Dynamical behavior of epidemiological models with nonlinear incidence rates'', Journal of Mathematical Biology 25.4 (1987): 359-380.


\bibitem{Janssen_Muller_Stenull2004}
H. K. Janssen, M. Müller, O. Stenull, ``Generalized epidemic process and tricritical dynamic percolation'', Physical Review E 70.2 (2004): 026114.

\bibitem{Dodds_Watts2005}
P. S. Dodds, D. J. Watts, ``A generalized model of social and biological contagion'', Journal of Theoretical Biology 232.4 (2005): 587-604.

\bibitem{Gross_Lima_Blasius2006}
T. Gross, C. D. D'Lima, B. Blasius, ``Epidemic dynamics on an adaptive network'', Physical Review Letters 96.20 (2006): 208701.


\bibitem{Bagnoli_Lio_Sguanci2007}
F. Bagnoli, P. Lio, L. Sguanci, ``Risk perception in epidemic modeling'', Physical Review E 76.6 (2007): 061904.


\bibitem{Bizhani_Paczuski_Grassberger2012}
G. Bizhani, M. Paczuski, P. Grassberger, ``Discontinuous percolation transitions in epidemic processes, surface depinning in random media, and Hamiltonian random graphs'', Physical Review E 86.1 (2012): 011128.

\bibitem{GomezGardenes2015}
J. G\'{o}mez-Garde\~{n}es \textit{et. al.}, ``Abrupt transitions from reinfections in social contagions'', Europhysics Letters 110.5 (2015): 58006.

\bibitem{Cai2015}
W. Cai \textit{et. al.}, "Avalanche outbreaks emerging in cooperative contagions." Nature Physics 11.11 (2015): 936-940.

\bibitem{Chae_Yook_Kim2015}
H. Chae, S. H. Yook, Y. Kim, ``Discontinuous phase transition in a core contact process on complex networks'', New Journal of Physics 17.2 (2015): 023039.

\bibitem{GomezGardenes2016}
J. G\'{o}mez-Garde\~{n}es \textit{et. al.}, ``Explosive contagion in networks'', Scientific Reports 6 (2016) 19767.

\bibitem{Liu2017}
Q. H. Liu \textit{et. al.}, ``Explosive spreading on complex networks: The role of synergy'', Physical Review E 95 (2017) 042320.

\bibitem{VelasquezRojas_Vazquez2017}
F. Vel\'{a}squez-Rojas, F. Vazquez, ``Interacting opinion and disease dynamics in multiplex networks: Discontinuous phase transition and nonmonotonic consensus times'', Physical Review E 95 (2017) 052315.

\bibitem{cui2017}
Pengi-Bi Cui, F. Colaiori, C. Castellano, ``Mutually cooperative epidemics on power-law networks'', Physical Review E 96 (2017) 022301.

\bibitem{KribsZaleta_VelascoHernandez2000}
C. M. Kribs-Zaleta, J. X. Velasco-Hernandez, ``A simple vaccination model with multiple endemic states'', Mathematical biosciences 164.2 (2000): 183-201.

\bibitem{Dorell2013}
C. Dorell \textit{et. al.}, ``Factors that influence parental vaccination decisions for adolescents, 13 to 17 years old national immunization survey–teen, 2010'', Clinical pediatrics 52.2 (2013): 162-170.


\end{thebibliography}

\end{document}